\documentclass[conference,compsoc]{IEEEtran}

\ifCLASSOPTIONcompsoc
  \usepackage[nocompress]{cite}
\else
  \usepackage{cite}
\fi

\ifCLASSINFOpdf
\else
\fi

\usepackage{graphicx}
\usepackage{booktabs}
\usepackage{array}
\usepackage{enumitem}%
\usepackage[hyphens]{url}
\usepackage{pifont}%
\usepackage{tikz}%
\usetikzlibrary{arrows.meta,fit,backgrounds,positioning}
\usepackage{algorithm}%
\usepackage[noend]{algpseudocode}
\newcommand{\BgImpactClasses}{12}
\newcommand{\BgImpactSubtypes}{36}
\newcommand{\PlDiffsRead}{300}
\newcommand{\PlLabels}{20}

\newcommand{\ChRecords}{3,816}
\newcommand{\ChFunctions}{7,534}
\newcommand{\ChRepos}{348}
\newcommand{\ChPartial}{1,633}
\newcommand{\ChLabeled}{3,748}
\newcommand{\ChUnknownLabels}{68}
\newcommand{\ChFineQueries}{143}
\newcommand{\ChFinePairs}{1,686}
\newcommand{\ChMixQueries}{84}
\newcommand{\ChMixPairs}{447}
\newcommand{\ChCRCapture}{13.3}
\newcommand{\ChCRMix}{7.1}
\newcommand{\ChCRCategoryPairAgreement}{12.2}
\newcommand{\ChCRCategoryCoverage}{80.5}
\newcommand{\ChCRSil}{0.307}
\newcommand{\ChCRDBI}{0.884}
\newcommand{\ChJinaCapture}{20.0}
\newcommand{\ChJinaMix}{44.1}
\newcommand{\ChJinaCategoryPairAgreement}{14.0}
\newcommand{\ChJinaCategoryCoverage}{84.7}
\newcommand{\ChJinaSil}{0.055}
\newcommand{\ChJinaDBI}{2.084}
\newcommand{\ChSCORCapture}{16.0}
\newcommand{\ChSCORMix}{11.9}
\newcommand{\ChSCORCategoryPairAgreement}{8.4}
\newcommand{\ChSCORCategoryCoverage}{77.8}
\newcommand{\ChSCORSil}{0.124}
\newcommand{\ChSCORDBI}{1.640}
\newcommand{\ChLTFCapture}{15.0}
\newcommand{\ChLTFMix}{13.5}
\newcommand{\ChLTFCategoryPairAgreement}{10.3}
\newcommand{\ChLTFCategoryCoverage}{78.6}
\newcommand{\ChLTFSil}{0.282}
\newcommand{\ChLTFDBI}{1.361}

\newcommand{\ChCatCheckShare}{53.9}
\newcommand{\ChCatBoundShare}{5.2}
\newcommand{\ChCatCapacityShare}{2.8}
\newcommand{\ChCatNumericShare}{2.3}
\newcommand{\ChCatSizeShare}{4.7}
\newcommand{\ChCatInitShare}{7.8}
\newcommand{\ChCatCleanupShare}{6.1}
\newcommand{\ChCatLifetimeShare}{7.9}
\newcommand{\ChCatFailureShare}{0.6}
\newcommand{\ChCatConcurrencyShare}{3.4}
\newcommand{\ChCatDataShare}{0.8}
\newcommand{\ChCatRestrictShare}{2.6}
\newcommand{\ChCatUnknownShare}{1.8}

\newcommand{\ChIxValidCVEs}{3,888}
\newcommand{\ChIxCoverageShare}{74.4}
\newcommand{\ChIxMemoryN}{975}
\newcommand{\ChIxMemoryShare}{25.6}
\newcommand{\ChIxStateN}{564}
\newcommand{\ChIxStateShare}{14.8}
\newcommand{\ChIxReferenceN}{387}
\newcommand{\ChIxReferenceShare}{10.1}

\newcommand{\IxChainMultiN}{198}
\newcommand{\IxChainLinkedN}{134}
\newcommand{\IxChainLinksN}{138}
\newcommand{\IxChainUpDownN}{111}
\newcommand{\IxChainNumMemN}{56}
\newcommand{\IxChainConStateN}{23}
\newcommand{\IxDosClasses}{12}
\newcommand{\IxCweTwentyClasses}{11}
\newcommand{\IxNoClassCVEs}{1,002}
\newcommand{\IxNoClassShareCVE}{25.8}
\newcommand{\IxChainLinkedShare}{67.7}
\newcommand{\IxChainUpDownShare}{80.4}
\newcommand{\IxHumanN}{100}
\newcommand{\IxHumanAgreeShare}{84.0}
\newcommand{\IxHumanMicroFone}{0.902}
\newcommand{\ChFiveBlocks}{283}
\newcommand{\ChFiveTestable}{198}
\newcommand{\ChFiveMain}{22}
\newcommand{\ChFivePointerPc}{55.0}
\newcommand{\ChFivePointerOther}{2.9}
\newcommand{\ChFiveCleanupPc}{67.8}
\newcommand{\ChFiveCleanupOther}{2.2}
\newcommand{\ChSixTargets}{291}
\newcommand{\ChSixRepos}{75}
\newcommand{\ChSixSourceGroups}{62}
\newcommand{\ChSixEasy}{111}
\newcommand{\ChSixMedium}{107}
\newcommand{\ChSixHard}{73}
\newcommand{\ChSixPoolPatches}{5,642}
\newcommand{\ChSixMeanZero}{37.9}
\newcommand{\ChSixMeanRandom}{36.9}
\newcommand{\ChSixMeanCluster}{49.2}
\newcommand{\ChSixMeanImpact}{48.5}
\newcommand{\ChSixMeanOracle}{56.9}
\newcommand{\ChSixMeanClusterGain}{12.3}
\newcommand{\ChSixMeanImpactGain}{11.6}
\newcommand{\ChSixEasyClusterGain}{20.0}
\newcommand{\ChSixMediumClusterGain}{10.3}
\newcommand{\ChSixHardClusterGain}{3.8}
\newcommand{\ChSixEasyImpactGain}{16.4}
\newcommand{\ChSixMediumImpactGain}{11.2}
\newcommand{\ChSixHardImpactGain}{4.8}
\newcommand{\ChSixMediumOracleGain}{24.1}
\newcommand{\ChSixHardOracleGain}{10.3}
\newcommand{\ChSixMeanClusterHyp}{41.6}
\newcommand{\ChSixMeanImpactHyp}{43.2}
\newcommand{\ChSixQueryGainCluster}{$+7.4$}
\newcommand{\ChSixQueryGainImpact}{$+4.8$}
\newcommand{\ChSixFilterHyp}{$+1.7$}
\newcommand{\ChSixFilterDev}{$-0.8$}
\newcommand{\ChSixFilterDiff}{$-2.4$}
\newcommand{\ChSixXplcImpactDevGain}{$+8.3$}
\newcommand{\ChSixXplcImpactDevChange}{$-3.3$}
\newcommand{\ChSixLineageOracleN}{257}
\newcommand{\ChSixXplcImpactGain}{$+4.4$}
\newcommand{\ChSixXplcImpactChange}{$-1.9$}
\newcommand{\ChSixJinaImpactHyp}{$+2.0$}
\newcommand{\ChSixJinaCluster}{$+3.3$}
\newcommand{\ChSixJinaImpact}{$+1.9$}
\newcommand{\ChSixJinaClusterHyp}{$+0.9$}
\newcommand{\ChSixJinaFilterHyp}{$+2.8$}
\newcommand{\RepairHumanN}{60}
\newcommand{\RepairHumanPathN}{55}
\newcommand{\RepairHumanPathShare}{91.7}
\newcommand{\RepairHumanConfirmedN}{38}
\newcommand{\RepairHumanPartialN}{6}
\newcommand{\RepairHumanIncorrectN}{2}
\newcommand{\RepairHumanUnknownN}{14}
\newcommand{\RepairHumanUnknownPathN}{13}
\newcommand{\ChGalleryPatches}{6,173}
\newcommand{\ChUnifiedImpactPatches}{2,838}
\newcommand{\ChAssocPairedPatches}{2,791}

\usepackage{xcolor}
\usepackage{colortbl}%
\usepackage{multirow}
\usepackage{verbatim}
\definecolor{patchadded}{HTML}{E6FFEC}
\definecolor{patchdeleted}{HTML}{FFEBE9}

\newcommand{\PatchLineBackground}[2]{%
  \begingroup
  \setlength{\fboxsep}{0pt}%
  \noindent\rlap{\colorbox{#1}{\strut\hspace{\linewidth}}}#2\par
  \endgroup}

\ExplSyntaxOn
\cs_new_protected:Npn \PatchFormatLine #1
  {
    \tl_if_head_eq_charcode:nNTF {#1} +
      { \PatchLineBackground{patchadded}{#1} }
      {
        \tl_if_head_eq_charcode:nNTF {#1} -
          { \PatchLineBackground{patchdeleted}{#1} }
          { #1\par }
      }
  }
\ExplSyntaxOff

\makeatletter
\renewcommand{\verbatim@processline}{%
  \expandafter\PatchFormatLine\expandafter{\the\verbatim@line}}
\makeatother

\newcounter{casectr}
\newcommand{\numberedcase}[1]{\refstepcounter{casectr}\label{#1}\noindent\textbf{Case~\thecasectr.}}
\begin{document}
\newcommand{\preprinttitle}{Not Just Any Example: Security Requirements and Fix Patterns for Guiding LLM Vulnerability Repair}
\title{\preprinttitle}
\author{%
  \IEEEauthorblockN{Zichen Liu\textsuperscript{1}, Liangyi Huang\textsuperscript{1},
    Stanley Nguyen\textsuperscript{1}, Kha Tu Nguyen\textsuperscript{1},
    Xinda Wang\textsuperscript{2}, and Xusheng Xiao\textsuperscript{1}}
  \IEEEauthorblockA{\textsuperscript{1}Arizona State University
    \qquad \textsuperscript{2}The University of Texas at Dallas}
  \IEEEauthorblockA{\small\ttfamily
    zliu396@asu.edu,\enspace lhuan139@asu.edu,\enspace snguye61@asu.edu\\
    knguy170@asu.edu,\enspace xinda.wang@utdallas.edu,\enspace Xusheng.Xiao@asu.edu}}
\ifdefined\pdfinfo
  \pdfinfo{/Title (\preprinttitle)
    /Author (Zichen Liu, Liangyi Huang, Stanley Nguyen, Kha Tu Nguyen, Xinda Wang, Xusheng Xiao)}
\fi

\maketitle

\begin{abstract}
Software vulnerabilities leave systems open to attack, and their remediation is often delayed. Large language models (LLMs) have advanced automated patch generation, yet passing functionality tests does not ensure that a patch enforces the security requirement the vulnerability violated. Existing vulnerability repair techniques rely on curated patterns or templates with limited coverage, leaving unclear which repair strategies carry over across vulnerabilities and how they can steer LLMs.

We present a large-scale empirical study that relates recurring fix patterns to the security impacts of vulnerabilities and examines how this knowledge can guide LLM-based repair. Building on a manual study of security patches, we cluster \ChRecords{} security patches from \ChRepos{} open-source repositories and analyze the edits and their context to derive concrete fix patterns, ranging from local safeguards to coordinated changes across multiple locations. We define \BgImpactClasses{} security impact classes by the security requirements that vulnerabilities violate and map them to repair categories and fix patterns, revealing characteristic repairs, alternative strategies within a class, and patterns shared across classes. We then use these patterns and impact--pattern associations to select repair examples for four LLMs on repair tasks of varying complexity. Guided example selection raises the four-model mean pass@3 from \ChSixMeanZero{}\% under zero-shot prompting to up to \ChSixMeanCluster{}\%, whereas randomly selected examples reach only \ChSixMeanRandom{}\%. Our findings connect vulnerability characterization with patch generation and support organizing repair knowledge around security requirements.
\end{abstract}

\IEEEpeerreviewmaketitle

\section{Introduction}
\label{sec:intro}

Software vulnerabilities allow attackers to disrupt or manipulate system behavior, making timely patching essential.
Yet operational constraints, staffing shortages, and competing priorities often delay remediation~\cite{dissanayake2022patchmanagement}, prolonging exposure to attacks~\cite{durumeric2014matter,li2017security}.
Reducing the effort required to develop fixes is one way to ease this burden, motivating automated program repair through test-driven patch search and learned transformations~\cite{weimer2009automatically,genprog2012,long2016angelic}.
For security vulnerabilities, however, passing functional tests is insufficient: generated patches can overfit the available tests without resolving the underlying security issue~\cite{qi2015analysis,smith2015is}.
Effective vulnerability patching must also restore violated security requirements, such as input validity, memory bounds and access control, while preserving intended functionality.
To address these requirements, vulnerability-specific repair approaches encode strategies in curated patterns and templates, but their coverage limits generalization to unseen vulnerability types and repair contexts~\cite{fang2024vfix,ma2017vurle}.
Recent advances in large language models (LLMs) offer an opportunity to broaden automated repair beyond these predefined strategies.
LLM-based patch generation substantially outperforms traditional and earlier learning-based techniques on functional-bug benchmarks~\cite{xia2023automated}, and LLM-based agents further extend repair to real-world software issues~\cite{li2025patchpilot,xia2025agentless}.

Despite these gains in functional-bug repair, where LLMs reason about functional requirements and code structure with feedback from tests, generating effective vulnerability patches remains challenging.
Security requirements additionally constrain behavior under adversarial inputs and execution conditions that functional tests often do not cover~\cite{fang2024vfix}.
Vulnerability repair must therefore prevent inputs from triggering exploitable behavior while preserving the program's intended functionality, requiring strategies that go beyond satisfying functional tests.
For example, a message parser may pass functional tests on well-formed messages yet copy an attacker-controlled number of bytes beyond a destination buffer.
A patch that checks only whether the destination pointer is non-null leaves this exploit possible.
An appropriate repair must validate the copy length against the destination capacity before copying, rejecting unsafe inputs while continuing to process valid messages.
This distinction illustrates why code-generation ability alone is insufficient: effective vulnerability patching also requires knowledge of which repair strategy restores the violated security requirement in the target context.
Making this knowledge explicit and accessible could help LLMs translate their generation capabilities into more reliable security patches.

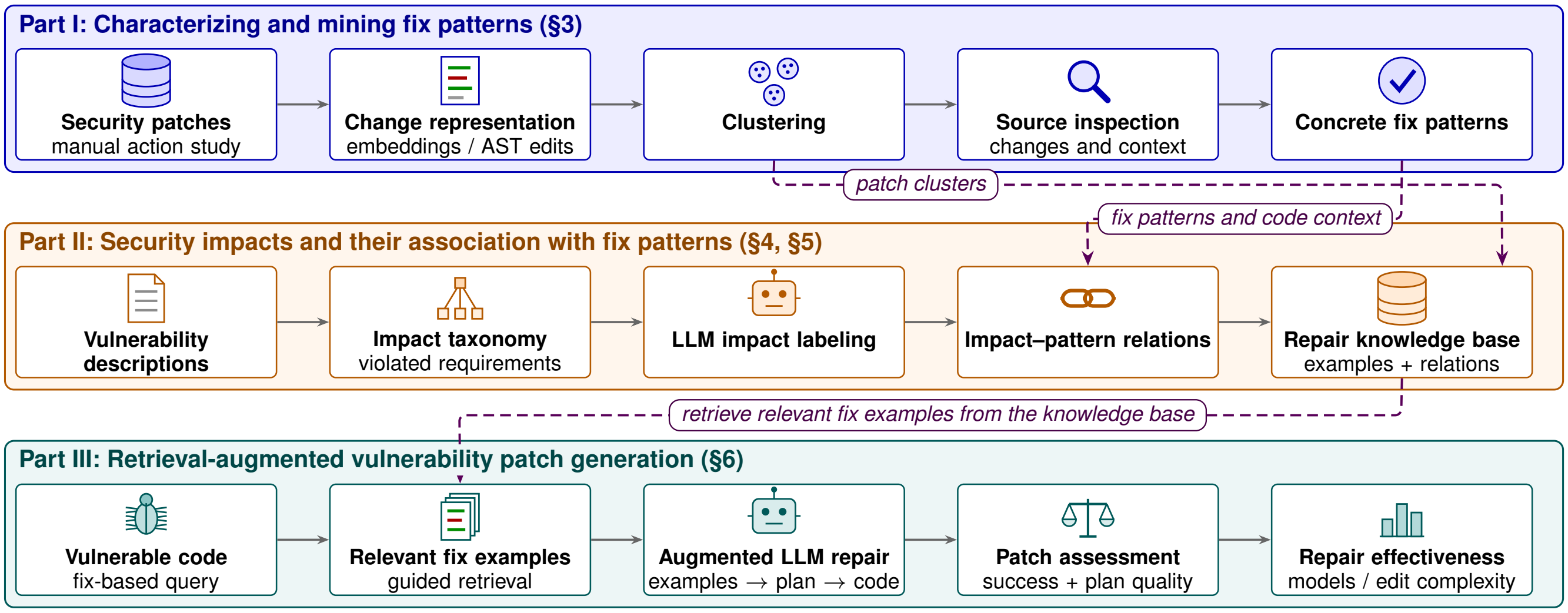
\begin{figure*}[t]
\centering
\resizebox{\textwidth}{!}{\tikzset{
  ic/.style={line width=0.55pt, draw=#1!70!black},
  pics/db/.style={code={
    \draw[ic=#1, fill=#1!15] (-.28,.22) -- (-.28,-.22) arc[start angle=180,end angle=360,x radius=.28,y radius=.09] -- (.28,.22);
    \draw[ic=#1, fill=#1!30] (0,.22) ellipse[x radius=.28,y radius=.09];
    \draw[ic=#1] (-.28,.02) arc[start angle=180,end angle=360,x radius=.28,y radius=.09];
    \draw[ic=#1] (-.28,-.10) arc[start angle=180,end angle=360,x radius=.28,y radius=.09];}},
  pics/diff/.style={code={
    \draw[ic=#1, fill=white] (-.22,-.28) rectangle (.22,.28);
    \draw[line width=1.1pt, green!55!black] (-.14,.15) -- (.12,.15);
    \draw[line width=1.1pt, red!65!black] (-.14,.03) -- (.08,.03);
    \draw[line width=1.1pt, green!55!black] (-.14,-.09) -- (.14,-.09);
    \draw[line width=1.1pt, black!40] (-.14,-.20) -- (.04,-.20);}},
  pics/clusters/.style={code={
    \foreach \cx/\cy in {-.17/.10, .16/.14, 0/-.17} {
      \draw[ic=#1, fill=#1!10] (\cx,\cy) circle (.12);
      \foreach \dx/\dy in {-.04/.03, .04/.04, 0/-.04} \fill[#1!70!black] (\cx+\dx,\cy+\dy) circle (.022);}}},
  pics/robot/.style={code={
    \draw[ic=#1] (0,.2) -- (0,.3); \fill[#1!70!black] (0,.31) circle (.035);
    \draw[ic=#1, fill=#1!15, rounded corners=2pt] (-.25,-.2) rectangle (.25,.2);
    \fill[#1!70!black] (-.1,.05) circle (.045) (.1,.05) circle (.045);
    \draw[ic=#1] (-.1,-.09) -- (.1,-.09);
    \draw[ic=#1] (-.25,0) -- (-.31,0) (.25,0) -- (.31,0);}},
  pics/check/.style={code={
    \draw[ic=#1, fill=#1!12] (0,0) circle (.27);
    \draw[line width=1.4pt, #1!70!black] (-.13,0) -- (-.03,-.11) -- (.15,.12);}},
  pics/doc/.style={code={
    \draw[ic=#1, fill=white] (-.21,-.28) -- (-.21,.28) -- (.10,.28) -- (.21,.17) -- (.21,-.28) -- cycle;
    \draw[ic=#1] (.10,.28) -- (.10,.17) -- (.21,.17);
    \foreach \y in {.08,-.03,-.14} \draw[black!45, line width=0.8pt] (-.13,\y) -- (.13,\y);}},
  pics/tree/.style={code={
    \draw[ic=#1] (0,.18) -- (-.2,-.15) (0,.18) -- (0,-.15) (0,.18) -- (.2,-.15);
    \draw[ic=#1, fill=#1!30] (-.06,.12) rectangle (.06,.24);
    \foreach \x in {-.2,0,.2} \draw[ic=#1, fill=#1!12] (\x-.06,-.24) rectangle (\x+.06,-.12);}},
  pics/link/.style={code={
    \draw[ic=#1, line width=1.2pt, rounded corners=3pt] (-.30,-.09) rectangle (.04,.09);
    \draw[ic=#1, line width=1.2pt, rounded corners=3pt] (-.04,-.09) rectangle (.30,.09);}},
  pics/magnifier/.style={code={
    \draw[ic=#1, fill=#1!10, line width=1pt] (-.05,.05) circle (.17);
    \draw[#1!70!black, line width=2pt, line cap=round] (.08,-.08) -- (.23,-.23);}},
  pics/bug/.style={code={
    \foreach \y in {.07,-.03,-.13} \draw[ic=#1] (-.24,\y-.03) -- (.24,\y+.03) (-.24,\y+.03) -- (.24,\y-.03);
    \draw[ic=#1, fill=#1!25] (0,-.03) ellipse[x radius=.13,y radius=.18];
    \draw[ic=#1, fill=#1!45] (0,.19) circle (.07);
    \draw[ic=#1] (0,.15) -- (0,-.2);}},
  pics/bulb/.style={code={
    \draw[ic=#1, fill=yellow!35] (0,.07) circle (.17);
    \draw[ic=#1, fill=#1!20] (-.08,-.12) rectangle (.08,-.22);
    \draw[ic=#1] (-.06,-.25) -- (.06,-.25);}},
  pics/scales/.style={code={
    \draw[ic=#1, line width=0.8pt] (0,-.24) -- (0,.2) (-.22,.14) -- (.22,.14) (-.1,-.24) -- (.1,-.24);
    \draw[ic=#1, fill=#1!15] (-.30,-.04) arc[start angle=180,end angle=360,radius=.08] -- cycle;
    \draw[ic=#1, fill=#1!15] (.14,-.04) arc[start angle=180,end angle=360,radius=.08] -- cycle;
    \draw[ic=#1] (-.22,.14) -- (-.30,-.04) (-.22,.14) -- (-.14,-.04) (.22,.14) -- (.14,-.04) (.22,.14) -- (.30,-.04);}},
  pics/stack/.style={code={
    \foreach \o in {.10,.05,0} \draw[ic=#1, fill=white] (-.22+\o,-.28+\o) rectangle (.12+\o,.16+\o);
    \draw[line width=1.0pt, green!55!black] (-.15,.06) -- (.05,.06);
    \draw[line width=1.0pt, red!65!black] (-.15,-.05) -- (.02,-.05);
    \draw[line width=1.0pt, green!55!black] (-.15,-.16) -- (.06,-.16);}},
  pics/chart/.style={code={
    \draw[ic=#1] (-.26,-.24) -- (.26,-.24);
    \foreach \x/\h in {-.17/.2, 0/.38, .17/.28} \draw[ic=#1, fill=#1!30] (\x-.06,-.24) rectangle (\x+.06,-.24+\h);}},
}
\begin{tikzpicture}[
  font=\sffamily\scriptsize,
  >={Stealth[length=4pt]},
  lane/.style={draw=#1!70!black, fill=#1!7, rounded corners=3pt, line width=0.6pt},
  step/.style={draw=#1!70!black, fill=white, rounded corners=2pt, line width=0.5pt,
               minimum width=3.05cm, minimum height=1.3cm, anchor=north west},
  txt/.style={font=\sffamily\scriptsize, align=center, text width=2.95cm, anchor=north, inner sep=0pt},
  head/.style={font=\sffamily\bfseries\footnotesize, text=#1!55!black, anchor=north west, inner sep=1pt, fill=#1!7},
  flow/.style={->, line width=0.7pt, draw=black!60, rounded corners=2pt},
  feed/.style={-{Stealth[length=5pt,width=4pt]}, line width=0.75pt, draw=violet!70!black, dash pattern=on 3.6pt off 2pt, rounded corners=2pt},
  lbl/.style={font=\sffamily\itshape\scriptsize, text=violet!55!black, fill=white, draw=violet!55!black, line width=0.45pt,
              rounded corners=3.5pt, inner xsep=4pt, inner ysep=1.8pt},
]
\newcommand{\stp}[6]{%
  \node[step=#2] (#1) at (#3,#4) {};
  \pic at ([yshift=-0.38cm]#1.north) {#5=#2};
  \node[txt] at ([yshift=-0.76cm]#1.north) {#6};}
\draw[lane=blue]   (-0.12,0)     rectangle (18.12,-1.95);
\draw[lane=orange] (-0.12,-2.55) rectangle (18.12,-4.5);
\draw[lane=teal]   (-0.12,-5.1)  rectangle (18.12,-7.05);

\stp{a1}{blue}{0}{-0.5}{db}{\textbf{Security patches}\\manual action study}
\stp{a2}{blue}{3.675}{-0.5}{diff}{\textbf{Change representation}\\embeddings / AST edits}
\stp{a3}{blue}{7.35}{-0.5}{clusters}{\textbf{Clustering}}
\stp{a4}{blue}{11.025}{-0.5}{magnifier}{\textbf{Source inspection}\\changes and context}
\stp{a5}{blue}{14.7}{-0.5}{check}{\textbf{Concrete fix patterns}}
\foreach \u/\v in {a1/a2,a2/a3,a3/a4,a4/a5} \draw[flow] (\u.east) -- (\v.west);

\stp{b1}{orange}{0}{-3.05}{doc}{\textbf{Vulnerability descriptions}}
\stp{b2}{orange}{3.675}{-3.05}{tree}{\textbf{Impact taxonomy}\\violated requirements}
\stp{b3}{orange}{7.35}{-3.05}{robot}{\textbf{LLM impact labeling}}
\stp{b4}{orange}{11.025}{-3.05}{link}{\textbf{Impact--pattern relations}}
\stp{b5}{orange}{14.7}{-3.05}{db}{\textbf{Repair knowledge base}\\examples + relations}
\foreach \u/\v in {b1/b2,b2/b3,b3/b4} \draw[flow] (\u.east) -- (\v.west);
\draw[flow] (b4.east) -- (b5.west);

\stp{c1}{teal}{0}{-5.6}{bug}{\textbf{Vulnerable code}\\fix-based query}
\stp{c2}{teal}{3.675}{-5.6}{stack}{\textbf{Relevant fix examples}\\guided retrieval}
\stp{c3}{teal}{7.35}{-5.6}{robot}{\textbf{Augmented LLM repair}\\examples $\rightarrow$ plan $\rightarrow$ code}
\stp{c4}{teal}{11.025}{-5.6}{scales}{\textbf{Patch assessment}\\success + plan quality}
\stp{c5}{teal}{14.7}{-5.6}{chart}{\textbf{Repair effectiveness}\\models / edit complexity}
\foreach \u/\v in {c1/c2,c2/c3,c3/c4,c4/c5} \draw[flow] (\u.east) -- (\v.west);

\draw[feed] (a5.south) -- (16.225,-2.5) -- (12.55,-2.5) -- (b4.north);
\node[lbl] at (14.4,-2.5) {fix patterns and code context};
\draw[feed] (a3.south) -- (8.875,-2.1) -- (17.4,-2.1) -- ([xshift=1.175cm]b5.north);
\node[lbl] at (10.6,-2.1) {patch clusters};
\draw[feed] (b5.south) -- (16.225,-4.8) -- (5.2,-4.8) -- (c2.north);
\node[lbl] at (10.8,-4.8) {retrieve relevant fix examples from the knowledge base};

\node[head=blue]   at (0,-0.06)  {Part I: Characterizing and mining fix patterns (\S\ref{sec:patterns})};
\node[head=orange] at (0,-2.61)  {Part II: Security impacts and their association with fix patterns (\S\ref{sec:impact-axis}, \S\ref{sec:pattern-impact})};
\node[head=teal]   at (0,-5.16)  {Part III: Retrieval-augmented vulnerability patch generation (\S\ref{sec:repair-check})};
\end{tikzpicture}}
\caption{Study overview: careful inspection and clustering of security patches reveal fix patterns. Security-impact analysis relates these patterns to violated requirements, forming a repair knowledge base that supports retrieval-augmented patch generation with relevant fix examples. Evaluation compares guided retrieval with Zero-shot, Random and Oracle conditions across models and edit complexities.}
\label{fig:overview}
\end{figure*}

Our careful inspection of vulnerability patches across projects reveals a basis for organizing this repair knowledge: patches with different syntactic structures often implement similar repair strategies.
For example, an early return and a conditional wrapper can both prevent a copy from exceeding its destination capacity.
This shared structure motivates mining \emph{fix patterns}, which capture recurring relations between code changes, the operations they protect, and the conditions under which the repair applies.
We describe a vulnerability's \emph{security impact} through the security requirements its code violates, such as memory bounds or the validity of a reference.
Relating impacts to patterns connects what a repair must restore with how code changes restore it, while the target's context determines which strategy applies.
The potential of these relationships to organize repair knowledge and guide LLM patch generation remains underexplored.

While existing work mines repair patterns and incorporates vulnerability knowledge into patch generation, its focus on edit transformations, selected vulnerability classes, or CWE-based organization leaves a gap in understanding how repair strategies address violated security requirements across code contexts.
Pattern miners extract recurring transformations from patches, including fixes for static-analysis warnings~\cite{koyuncu2020fixminer,bader2019getafix,liu2017findbugs,yang2022pythonfix,rolim2018revisar}.
Repair approaches use curated patterns, templates and learned transformations to generate patches~\cite{fang2024vfix,ma2017vurle,harer2018learning}.
Impact-aware work summarizes security operations and patch patterns by impact~\cite{wu2020sid}, encodes expected repairs for memory bug types in an LLM prompt~\cite{li2026whatdotheyfix}, or organizes repair knowledge by CWE type~\cite{li2026vulkey}.
These perspectives do not yet provide a comprehensive, large-scale account of alternative repair strategies for the same security impact, strategies shared across impacts, and the code conditions that determine their applicability.
Without this account, it remains unclear how fix-pattern and security-impact knowledge can support retrieval-augmented generation by retrieving only fix examples relevant to the target vulnerability from a repair knowledge base.
We therefore conduct a large-scale study to develop an in-depth understanding of fix patterns, their associations with security impacts, and how this knowledge can improve LLMs' vulnerability patch generation.
The study aims to establish an empirical foundation for further research on integrating repair knowledge with LLM patch generation.

\noindent\textbf{Challenges.}
Conducting this study requires addressing three challenges in identifying fix patterns, relating them to security impacts, and evaluating their use in LLM patch generation.
\ding{182}~\emph{Recognizing repair mechanisms across code forms.}
The same edit can protect different operations, while different edits or coordinated changes can implement one pattern.
Representations and source analysis must capture the relationship between the modification, its context and the requirement restored.
\ding{183}~\emph{Characterizing the requirement a vulnerability violates.}
Descriptions combine implementation defects, violated requirements and attack consequences, and may report several successive violations.
A consistent impact taxonomy must distinguish these levels and retain the relationships among violations.
\ding{184}~\emph{Turning repair knowledge into useful guidance.}
A fix example must convey a repair relationship applicable to the target.
Evaluation must establish both the benefit of relevant fix examples and how well retrieval obtains such guidance across repairs of different complexity.

\noindent\textbf{Large-scale study.}
In our study, we first use ARVO~\cite{mei2026arvo} for a manual fix-action study and then use PatchDB, CVEfixes and MegaVul~\cite{wang2021patchdb,bhandari2021cvefixes,ni2024megavul} for large-scale patch analysis.
The study proceeds in three connected parts (Fig.~\ref{fig:overview}).

\noindent\emph{Part~I: Characterizing and mining fix patterns.}
We begin by examining the fix actions in \PlDiffsRead{} security patches and find recurring actions and combinations.
These observations motivate broad repair categories and the search for more specific patterns in \ChRecords{} patches from \ChRepos{} repositories.
We compare CodeReviewer and Jina code embeddings with AST-edit representations, assessing cluster geometry and the concentration of related repair categories and same-pattern examples (Section~\ref{sec:patterns}).
Source comparison within the clusters identifies concrete patterns for initializing output before export, bounding recursive processing, preserving string termination, and other recurring repairs.
These concrete patterns provide interpretable evidence of the repair mechanisms reflected in the clusters, complementing the quantitative evaluation of patch grouping.

\noindent\emph{Part~II: Classifying security impacts and relating them to fix patterns.}
Existing impact-aware studies often focus on selected vulnerability types, leaving the broader spectrum of security impacts incompletely characterized~\cite{wu2020sid,li2026whatdotheyfix}.
\textbf{We define a more comprehensive taxonomy of \BgImpactClasses{} security impact classes based on violated security requirements, drawing on studies of concurrency, algorithmic complexity, input interpretation and certificate validation, together with cryptographic specifications}~\cite{lu2008concurrency,crosby2003complexity,sassaman2013langsec,georgiev2012dangerous,mcgrew2008aead}.
We conduct a large-scale security-impact study on \ChRecords{} patches from \ChRepos{} diverse open-source projects collected from PatchDB, CVEfixes and MegaVul, uncovering a wide spectrum of violations across memory safety, concurrency, resource management, trust decisions and data protection~\cite{wang2021patchdb,bhandari2021cvefixes,ni2024megavul}.
We label vulnerability descriptions separately from the code-based characterization of repairs (Section~\ref{sec:impact-axis}).
The taxonomy captures individual violations and directed chains. The results also show why a consequence or a broad CWE label does not uniquely identify the requirement to restore.
We identify \ChFiveMain{} characteristic associations between impact classes and broad repair categories, then examine clustered source examples to explain concrete repair choices (Section~\ref{sec:pattern-impact}).
For memory-bounds, the code's supported input domain and lookup logic distinguish rejecting a request, increasing capacity and correcting an index calculation.
Impact thus narrows the candidate approaches, while code context explains their applicability.

\noindent\emph{Part~III: Retrieval-augmented LLM repair with relevant fix examples.}
We retrieve fix examples from the repair knowledge base to guide four LLMs through the same plan-then-code workflow (Section~\ref{sec:repair-check}).
Cluster-guided uses patch clusters to find similar code changes, while Impact-guided selects examples from repair categories associated with the target's security impact.
We compare both methods with Zero-shot, Random and an Oracle based on reviewed mechanism matches, evaluating repair success and plan quality on \ChSixTargets{} targets stratified by edit complexity.

Our evaluation results show that \textbf{choosing appropriate fix examples substantially improves LLM patch generation}: Cluster-guided and Impact-guided retrieval raise four-model mean pass@3 from \ChSixMeanZero{}\% with Zero-shot to \ChSixMeanCluster{}\% and \ChSixMeanImpact{}\%, respectively, while reviewed pattern matches reach \ChSixMeanOracle{}\%.
\textbf{The gains are particularly large for the open-weight models Qwen3.8-27B and GLM-4.5-Air.}
Relative to Random, Cluster-guided retrieval improves their pass@3 by 42.2\% and 80.3\%, respectively, while Impact-guided retrieval yields improvements of 39.2\% and 87.1\%.
\textbf{Relevant examples also enable some open-weight models to match or exceed GPT-6 Luna's observed repair rate on easy targets}: Cluster-guided retrieval yields 74.8\% pass@3 for Qwen3.8-27B and 75.7\% for gpt-oss-120b, compared with 73.9\% for GPT-6 Luna.
On medium targets, Impact-guided retrieval brings gpt-oss-120b to 47.7\%, close to GPT-6 Luna's 48.6\% under the same method.
These gains suggest that relevant examples supply vulnerability-specific repair knowledge that these models do not reliably recover from the target code alone, helping them identify the violated requirement and an applicable repair strategy.
\textbf{This finding highlights the potential of retrieval-augmented generation to strengthen open-weight models by supplementing their code-generation capabilities with relevant security-repair knowledge.}
However, not every example helps: with the same number of examples, random selection lowers observed repair rates for three of the four generators, yielding a mean pass@3 of \ChSixMeanRandom{}\%.
\textbf{These results highlight the need to retrieve fix examples whose repair strategies apply to the target vulnerability, so that generation draws on relevant knowledge from the repair knowledge base.}
Analysis of the generated plans further shows that relevant examples improve problem identification and repair-plan completeness, helping explain their benefit to patch generation.
Using the generator's own hypothetical fix as the retrieval query retains part of the improvement, suggesting that this knowledge can also help when the developer fix is unavailable.
Improvements nevertheless narrow on harder targets, where reviewed pattern matches retain a larger advantage, motivating further work on retrieving and applying suitable repair strategies.
By characterizing this knowledge and examining its use in patch generation, our study aims to inspire further research on integrating fix patterns and security impacts with LLMs for more effective vulnerability repair.

\section{Key Concepts}
\label{sec:background}
\label{sec:overview}

\label{sec:Ov-objects}

\noindent\textbf{Security impact and security requirements.}
A \emph{security requirement} is a condition that an operation must satisfy for the program to stay secure: a write stays inside its object, a reference has a valid target, an access carries the required permission.
We characterize security impact by the security requirements that vulnerable code violates~\cite{wu2020sid}.
A CVE description may state the violated requirement alongside the \emph{root cause} (the defect in the code) and the \emph{consequence} (what an attacker gains). Fig.~\ref{fig:running-example} marks all three for our running example.
We label the stated violation.
A root cause can already name the repair (in the example, a missing length check), so labeling it would encode the fix (Section~\ref{sec:Ov-separation}). By contrast, a consequence such as a crash can follow from many different violations.

An \emph{impact class} groups violations of related security requirements, such as memory bounds or object lifetime.
A vulnerability may involve more than one class.
The example in Fig.~\ref{fig:running-example} involves a write past a stack buffer's bounds, in the class \emph{memory-bounds}.

\newsavebox{\RunExImpactBox}
\newsavebox{\RunExFixBox}
\newlength{\RunExInnerW}
\begin{figure}[t]
\centering
\definecolor{impactbounds}{HTML}{E6F1E8}
\setlength{\RunExInnerW}{\dimexpr\linewidth-13pt\relax}
\newcommand{\RunExEvidence}[1]{{\setlength{\fboxsep}{1.2pt}\colorbox{impactbounds}{\vphantom{Ag}#1}}}
\newcommand{\RunExClass}[1]{\tikz[baseline=(c.base)]\node[rounded corners=2pt,fill=impactbounds,draw=green!40!black!70,line width=0.5pt,inner xsep=3pt,inner ysep=1.5pt,font=\itshape](c){#1};}
\begin{lrbox}{\RunExImpactBox}%
\begin{minipage}{\RunExInnerW}
\footnotesize
\renewcommand{\arraystretch}{1.12}%
\begin{tabular}{@{}>{\raggedright\arraybackslash\itshape}p{1.52cm}@{\hspace{5pt}}>{\raggedright\arraybackslash}p{\dimexpr\linewidth-1.52cm-5pt\relax}@{}}
Root cause & ``not ensuring that the key signature length in a Tag 11 packet is compatible with the key signature buffer size'' \\
Requirement violation & \RunExEvidence{``Stack-based buffer overflow''} \\
Consequence & ``cause a denial of service (system crash) or possibly gain privileges'' \\[1pt]
\arrayrulecolor{orange!35}\midrule[0.4pt]\arrayrulecolor{black}
\upshape\bfseries Impact class & \RunExClass{memory-bounds} (a write exceeds the key signature buffer's capacity).
\end{tabular}
\end{minipage}%
\end{lrbox}%
\begin{lrbox}{\RunExFixBox}%
\begin{minipage}{\RunExInnerW}
\begingroup\scriptsize
\input{generated/tables/Ex_running_code}
\endgroup
\vspace{1pt}{\color{blue!30}\hrule height 0.4pt}\vspace{4pt}
\footnotesize
\noindent\textbf{Pattern: Reject copies that exceed the destination buffer's capacity.}\par\smallskip
The added check exits before \texttt{memcpy} when the copy length exceeds the available capacity.
\end{minipage}%
\end{lrbox}%
\begin{tikzpicture}[
  rxhead/.style={anchor=south west, inner xsep=6pt, inner ysep=2.4pt, text width=\RunExInnerW, align=left,
                 font=\sffamily\footnotesize},
  rxbody/.style={anchor=north west, inner xsep=6pt, inner ysep=4pt},
]
\newcommand{\RunExFrame}[3]{%
  \begin{scope}[on background layer]
    \clip[rounded corners=3pt] (#2.north west) rectangle (#3.south east);
    \fill[#1!8] (#2.north west) rectangle (#2.south east);
  \end{scope}
  \draw[#1!45, line width=0.4pt] (#2.south west) -- (#2.south east);
  \draw[#1!65!black, line width=0.5pt, rounded corners=3pt] (#2.north west) rectangle (#3.south east);}
\node[rxbody] (ib) at (0,0) {\usebox{\RunExImpactBox}};
\node[rxhead] (ih) at (ib.north west)
  {{\bfseries\color{orange!60!black}Security impact}\enspace{\color{black!55}CVE description only}};
\RunExFrame{orange}{ih}{ib}
\node[rxhead, anchor=north west] (fh) at ([yshift=-6pt]ib.south west)
  {{\bfseries\color{blue!60!black}Fix pattern}\enspace{\color{black!55}code only}};
\node[rxbody] (fb) at (fh.south west) {\usebox{\RunExFixBox}};
\RunExFrame{blue}{fh}{fb}
\end{tikzpicture}
\caption{Security impact and fix pattern for a Linux eCryptfs stack-buffer overflow. \texttt{+} marks additions, and \texttt{...} marks omissions.}
\label{fig:running-example}
\end{figure}

\noindent\textbf{Fix patterns.}
A \emph{fix pattern} (or \emph{pattern}) is a recurring way of changing code to satisfy a security requirement.
Examples include rejecting a copy that exceeds its destination capacity (Fig.~\ref{fig:running-example}) and initializing an output structure before exposing its contents.
A pattern captures how the changes work with the surrounding code, so different edits and implementations can realize the same pattern.
A patch may implement several patterns, and a pattern may require coordinated changes across functions or files (Section~\ref{sec:Pt-concepts}).

\noindent\textbf{Other Important Concepts.}
A \emph{patch sample} is a dataset sample containing the available diff and recovered before-and-after code. Our fix-pattern study represents each sample as one clustering point.
A sample may contain changes to several functions or files and may implement several fix patterns.
Depending on the representation, we encode individual function changes or assemble code fragments from the patch. These input-construction choices do not define the boundaries of a fix pattern (Section~\ref{sec:Pt-representations}).
A \emph{cluster} groups patch samples with similar representations, which we inspect for shared fix patterns.
A \emph{fix example} presents pre-fix code and its corresponding patch to illustrate a repair. The repair experiments use function-level examples (Section~\ref{sec:Rc-setup}).
A CVE description is the unit of impact labeling and is counted once even when several patches address the CVE.
Patch-level analyses associate each patch with the impact classes of its linked CVEs, retaining several classes when distinct violations are stated.

\section{Part~I: Mining Fix Patterns}
\label{sec:patterns}

Our study first examines whether code representations group security patches by recurring repair mechanisms, combining a manual fix-action study, clustering evaluation and source-based pattern analysis.

\noindent\textbf{Datasets.}\label{sec:Ov-data} We collect C/C++ vulnerability patches from PatchDB~\cite{wang2021patchdb}, CVEfixes~\cite{bhandari2021cvefixes} and MegaVul~\cite{ni2024megavul}. ARVO~\cite{mei2026arvo} supplies patches for the manual study (Section~\ref{sec:Pt-manual}).
Each patch sample records its repository and fixing commit, linked CVE identifiers and available descriptions, the diff, and recovered before-and-after function code.
We consolidate records for the same fix using repository and commit identifiers, and link equivalent backports into repair families (Appendix~\ref{app:design}).

The clustering study uses \ChRecords{} patch samples from \ChRepos{} repositories.
Our security-impact and vulnerability-repair studies specify how they reuse this corpus and distinguish their analysis and repair-target populations (Sections~\ref{sec:impact-axis} and~\ref{sec:repair-check}).

\subsection{From Fix Actions to Fix Patterns}
\label{sec:Pt-concepts}
\label{sec:Pt-manual}
\label{sec:Pt-pilot}

\noindent\textbf{Manual study of fix actions.}
We first examined which code changes recur in security patches and how they combine.
Two authors inspected \PlDiffsRead{} ARVO patch diffs~\cite{mei2026arvo}, deduplicated by fixing commit, and labeled the changed lines with the actions they perform.
They started from fix actions described in a study of Python bug fixes~\cite{yang2022pythonfix}, such as changing the number or values of call arguments, and added labels when existing ones did not fit.
The resulting list contains \PlLabels{} action labels, including variants of the same action.
Table~\ref{tab:Pt-manual} summarizes the frequently labeled actions and shows the prominence of changes to precondition checks.
Most inspected patches received several labels, and some combined changes at non-contiguous locations or across files.
These observations motivate looking for recurring fix patterns implemented through individual actions or combinations of actions.

\newcommand{\PlApproxPreAdd}{210}
\newcommand{\PlApproxPreAddDel}{200}
\newcommand{\PlApproxPreChange}{185}
\newcommand{\PlApproxPreRemove}{185}
\newcommand{\PlApproxCallChange}{135}
\newcommand{\PlApproxConditionChange}{135}
\newcommand{\PlApproxPostAdd}{115}
\newcommand{\PlApproxAssignment}{115}
\newcommand{\PlApproxPostChange}{115}
\newcommand{\PlSharePreAdd}{15.1}
\newcommand{\PlSharePreAddDel}{14.3}
\newcommand{\PlSharePreChange}{13.3}
\newcommand{\PlSharePreRemove}{13.3}
\newcommand{\PlShareCallChange}{9.7}
\newcommand{\PlShareConditionChange}{9.7}
\newcommand{\PlSharePostAdd}{8.2}
\newcommand{\PlShareAssignment}{8.2}
\newcommand{\PlSharePostChange}{8.2}

\begin{table}[t]
\centering
\footnotesize
\setlength{\tabcolsep}{3pt}
\renewcommand{\arraystretch}{1.05}
\caption{Frequent fix actions with approximate counts and shares within the displayed rows.}
\label{tab:Pt-manual}
\begin{tabular*}{.94\columnwidth}{@{}l@{\extracolsep{\fill}}rr@{}}
\toprule
\textbf{Fix action} & \textbf{Approx. count} & \textbf{Share (\%)} \\
\midrule
Add precondition check & \PlApproxPreAdd{} & \PlSharePreAdd{} \\
Add or delete precondition check & \PlApproxPreAddDel{} & \PlSharePreAddDel{} \\
Change precondition check & \PlApproxPreChange{} & \PlSharePreChange{} \\
Remove precondition check & \PlApproxPreRemove{} & \PlSharePreRemove{} \\
Change method call & \PlApproxCallChange{} & \PlShareCallChange{} \\
Change precondition & \PlApproxConditionChange{} & \PlShareConditionChange{} \\
Add postcondition check & \PlApproxPostAdd{} & \PlSharePostAdd{} \\
Change variable assignment & \PlApproxAssignment{} & \PlShareAssignment{} \\
Change postcondition check & \PlApproxPostChange{} & \PlSharePostChange{} \\
\bottomrule
\end{tabular*}
\end{table}

\noindent\textbf{From edits to repair mechanisms.}
An edit's purpose depends on its context: an added condition may protect a read's extent, pointer validity or object state.
Prior work characterizes recurring edits~\cite{pan2009bugfixpatterns} and combines AST context with edit actions~\cite{koyuncu2020fixminer}. We examine the repair relationships established by changes and surrounding code.
A \emph{fix pattern} identifies the affected data or operation, its required condition, and the changes that establish it, such as checking that enough input remains before a read.
Broad categories group repairs by their main approach, while concrete patterns capture cooperating changes across categories.
For example, preserving string termination can combine reserving a buffer byte, bounding input to the remaining space, and keeping the reserved byte zero.

\subsection{Representing and Clustering Patches}
\label{sec:Pt-representations}
\label{sec:Pt-discovery}

Each patch sample contributes one clustering point and may span several functions or files and implement multiple patterns.
The fixed cohort contains \ChRecords{} samples and \ChFunctions{} recovered modified functions from a gallery of \ChGalleryPatches{} patches, without a per-sample function cap. Available code need not reconstruct every commit edit (Appendix~\ref{app:clustering}).
We compare frozen CodeReviewer (CR)~\cite{li2022codereviewer} and Jina~\cite{kryvosheieva2025jina} embeddings with Diff-AST Centrality (Diff-AST) and Edit-Script TF-IDF (TF-IDF), producing one unit-length vector per sample.

\noindent\textbf{Code inputs.}
Both embedding methods retain changed lines and up to three aligned context lines on each side, marking omitted spans.
CR encodes separate before and after views of each recovered modified function, without function titles.
Jina encodes marked raw-patch fragments, including changes outside functions, with original identifiers and numbered file and fragment headers.
For parseable code, Jina extends fragments to their containing statements or control headers using existing source lines. When the input budget is exceeded, optional context is removed before substantive changes.

\noindent\textbf{CR: encode changes within functions.}
We mean-pool CodeReviewer's encoder token states for each before and after input, retaining its unchanged-line markers.
This readout and the following change composition are study choices.
Let $u_f$ and $v_f$ be the L2-normalized embeddings of function $f$'s before and after views, and let $\mathcal{N}$ normalize a vector to unit length.
We form
\begin{equation}
\begin{array}{rl}
d_f &= \mathcal{N}(v_f-u_f),\\
z_{\mathrm{CR}} &= \mathcal{N}\!\left(\frac{1}{m}\sum_{f=1}^{m}d_f\right),
\end{array}
\label{eq:Pt-cr-change}
\end{equation}
where $m$ is the number of retained function changes.
This composition emphasizes changes relative to local context and weights functions equally.

\noindent\textbf{Jina: encode the joint patch.}
For Jina Code Embeddings 1.5B, we prefix the marked diff with its code-passage instruction and use the final token's hidden state, encoding changes and context jointly.

Long inputs are encoded in non-overlapping chunks and combined by length-weighted averaging.

\noindent\textbf{AST-edit baselines.}
Both structural baselines use GumTree edits between paired functions~\cite{falleri2014gumtree}.
Diff-AST Centrality describes the edit graph using node-centrality features, adapting TreeCen's structural-summary approach~\cite{hu2022treecen}, with a fixed function-level scaler to standardize its features.
Edit-Script TF-IDF applies a fixed TF-IDF vocabulary to edit-action and AST-context tokens.
For each baseline, normalized function vectors are averaged within a patch.

\noindent\textbf{Dimensionality reduction and clustering.}
All representations use average-linkage hierarchical clustering with 64 groups and Euclidean working-space distances.
CR uses UMAP~\cite{mcinnes2018umap} to reduce vectors to 128 dimensions, whereas Jina and both structural baselines use PCA with 32 dimensions (Table~\ref{tab:Pt-pipelines}). Clustering uses the reduced vectors directly.

\begin{table}[t]
\centering
\footnotesize
\setlength{\tabcolsep}{3pt}
\caption{Patch-clustering pipelines. $K$: number of clusters.}
\label{tab:Pt-pipelines}
\begin{tabular*}{.92\columnwidth}{@{\extracolsep{\fill}}lllr@{}}
\toprule
Representation & Reduction & Clustering & $K$ \\
\midrule
\multicolumn{4}{@{}l}{\emph{Code embeddings}} \\
CR & UMAP, 128D & Average linkage & 64 \\
Jina & PCA, 32D & Average linkage & 64 \\
\midrule
\multicolumn{4}{@{}l}{\emph{AST-edit features}} \\
Diff-AST & PCA, 32D & Average linkage & 64 \\
TF-IDF & PCA, 32D & Average linkage & 64 \\
\bottomrule
\end{tabular*}
\end{table}

\subsection{Evaluating Patch Clusters}
\label{sec:Pt-quality}

Table~\ref{tab:Pt-quality} reports geometric and semantic quality.
Silhouette~\cite{rousseeuw1987silhouette} measures within-cluster cohesion relative to neighboring clusters ($\uparrow$), whereas Davies--Bouldin (DBI)~\cite{davies1979cluster} measures dispersion relative to cluster-center separation ($\downarrow$).
Both use the working space (Appendix~\ref{app:clustering}).

\begin{table}[t]
\centering
\footnotesize
\setlength{\tabcolsep}{3pt}
\caption{Cluster quality on \ChRecords{} patches. All semantic scores are percentages.}
\label{tab:Pt-quality}
\label{tab:Pt-semantic}
\begin{tabular*}{.92\columnwidth}{@{\extracolsep{\fill}}lrrrrrr@{}}
\toprule
Method & Sil.\ $\uparrow$ & DBI $\downarrow$ & $C\uparrow$ & $M\downarrow$ & $T_3\uparrow$ & $P_{\times}\uparrow$ \\
\midrule
\multicolumn{7}{@{}l}{\emph{Code embeddings}} \\
CR & \ChCRSil{} & \ChCRDBI{} & \ChCRCapture{} & \ChCRMix{} & \ChCRCategoryCoverage{} & \ChCRCategoryPairAgreement{} \\
Jina & \ChJinaSil{} & \ChJinaDBI{} & \ChJinaCapture{} & \ChJinaMix{} & \ChJinaCategoryCoverage{} & \ChJinaCategoryPairAgreement{} \\
\midrule
\multicolumn{7}{@{}l}{\emph{AST-edit features}} \\
Diff-AST & \ChSCORSil{} & \ChSCORDBI{} & \ChSCORCapture{} & \ChSCORMix{} & \ChSCORCategoryCoverage{} & \ChSCORCategoryPairAgreement{} \\
TF-IDF & \ChLTFSil{} & \ChLTFDBI{} & \ChLTFCapture{} & \ChLTFMix{} & \ChLTFCategoryCoverage{} & \ChLTFCategoryPairAgreement{} \\
\bottomrule
\end{tabular*}
\end{table}

\noindent\textbf{Broad fix-pattern categories.}
Prior work~\cite{pan2009bugfixpatterns,sobreira2018dissection,koyuncu2020fixminer} and our manual study inform twelve repair categories (Table~\ref{tab:Pt-categories}). Checks and bounds additionally distinguish the protected constraint.
Labels support evaluation and source inspection, not encoder training or clustering.

\begin{table}[t]
\centering
\footnotesize
\setlength{\tabcolsep}{3pt}
\caption{Broad fix-pattern categories and shares, as percentages of all \ChRecords{} patches.}
\label{tab:Pt-categories}
\begin{tabular}{@{}>{\raggedright\arraybackslash}p{0.32\columnwidth}>{\raggedright\arraybackslash}p{0.51\columnwidth}r@{}}
\toprule
Category & Typical change & Share \\
\midrule
Check and reject & Reject an operation when a required condition fails. & \ChCatCheckShare{}\% \\
Bound processing & Limit a length, count, value, or workload. & \ChCatBoundShare{}\% \\
Provide capacity & Allocate enough space for the required data. & \ChCatCapacityShare{}\% \\
Change numeric representation & Correct a value's type, width, or signedness. & \ChCatNumericShare{}\% \\
Correct size or position & Correct a length, offset, or index calculation. & \ChCatSizeShare{}\% \\
Initialize or clear state & Define memory or state before use or output. & \ChCatInitShare{}\% \\
Release and clean up & Correct missing or repeated resource releases. & \ChCatCleanupShare{}\% \\
Protect object lifetime & Keep objects valid while they are used. & \ChCatLifetimeShare{}\% \\
Propagate failure & Return errors that were previously lost. & \ChCatFailureShare{}\% \\
Coordinate concurrent access & Synchronize access to shared state. & \ChCatConcurrencyShare{}\% \\
Separate data from instructions & Keep input as data in commands, queries, or formats. & \ChCatDataShare{}\% \\
Restrict privileges or features & Remove unnecessary permissions or functionality. & \ChCatRestrictShare{}\% \\
\midrule
Undetermined & The repair category cannot be established. & \ChCatUnknownShare{}\% \\
\bottomrule
\end{tabular}
\end{table}

\noindent\textbf{Labeling protocol.}
We use LLMs to assign broad fix-pattern categories and manually audit 100 patches covering all twelve categories to assess category applicability and labeling agreement.
Code changes and context take precedence over auxiliary repair-related commit messages or CVE passages. Impact labels are withheld.
The reference assigns a primary category to \ChLabeled{} patches. Category scores exclude the \ChUnknownLabels{} unresolved cases.

\noindent\textbf{Fix-pattern-based cluster evaluation.}
$T_3$ measures labeled-patch coverage by each cluster's three largest categories. $P_{\times}$ measures primary-label agreement among cross-source-group pairs within a cluster, using 22 labels that distinguish protected constraints.
Partner capture $C$ and distractor mixing $M$ measure the fractions of known same-pattern partners and reviewed same-source-group distractors placed with a test patch, averaging equally over patches with references.
Higher $T_3$, $P_{\times}$ and $C$, and lower $M$, are better. Source-group rules, reference counts and formulas appear in Appendix~\ref{app:clustering}.

Both embedding methods exceed the AST-edit baselines on $T_3$ and $P_{\times}$.
CR also improves Silhouette, DBI and distractor mixing, while Jina captures more same-pattern partners.
We retain both for source analysis: CR groups more selectively, while Jina gathers more known implementations.

\subsection{Fix Patterns in Patch Clusters}
\label{sec:Pt-mining}
\label{sec:Pt-catalog}
\label{sec:Pt-cases}
\label{sec:Pt-limits}

\begin{figure}[t]
\centering
\includegraphics[width=\columnwidth]{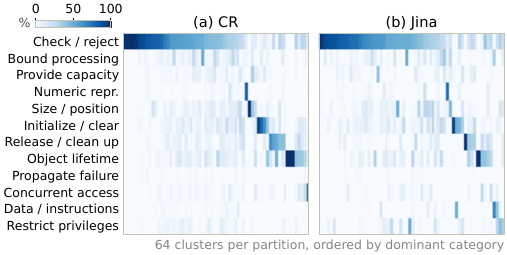}
\caption{Broad-category distributions across CR and Jina clusters. Each column is a cluster ordered by dominant category, and darker cells indicate larger within-cluster shares.}
\label{fig:Pt-category-map}
\end{figure}

To identify fix patterns, we inspect code changes and their surrounding context within the CR and Jina clusters.
We use the category distributions in Fig.~\ref{fig:Pt-category-map} to examine clusters dominated by initialization or checks as well as clusters containing several repair categories.
We propose candidate patterns from this inspection, then use LLM agents to refine their definitions against the source changes and check each cluster member for support.

\noindent\textbf{Finding 1: Patch clusters reflect shared repair mechanisms across different code forms.}
Source comparison reveals shared protections implemented through different code forms in CR and Jina clusters (Fig.~\ref{fig:Pt-cluster-cases}).

\emph{(1) Initialize output structures before exporting them.}
Patches clear entire output structures before field assignments, as in Linux cxgb3 (Fig.~\ref{fig:Pt-cluster-cases}a), or initialize reserved and padding fields to prevent exposure of undefined bytes.

\emph{(2) Carry and bound recursive depth or cost.}
Wireshark propagates depth across decoding calls (Fig.~\ref{fig:Pt-cluster-cases}b), while Libexif accumulates cost from entries at each level. Both stop processing when a bound is exceeded.

\emph{(3) Bound string input and preserve a terminator.}
ssdp-responder reserves a byte beyond bounded input in a zero-initialized buffer (Fig.~\ref{fig:Pt-cluster-cases}c), while GPAC explicitly zeros the reserved position. Both preserve string termination.

Table~\ref{tab:Pt-additional-patterns} summarizes further recurring patterns found in the clusters.

\begin{table}[t]
\centering
\footnotesize
\setlength{\tabcolsep}{3pt}
\renewcommand{\arraystretch}{1.08}
\caption{Additional fix patterns identified in patch clusters.}
\label{tab:Pt-additional-patterns}
\begin{tabular}{@{}>{\raggedright\arraybackslash}p{0.36\columnwidth}>{\raggedright\arraybackslash}p{0.59\columnwidth}@{}}
\toprule
Fix pattern & Core repair relation \\
\midrule
Check pointer chains & Verify that each pointer needed to reach a nested field is non-null before reading it. \\
Validate indices and lookup results & Reject out-of-range indices or failed lookup results before accessing a collection. \\
Limit access to remaining space & Check or cap a read or copy length using the bytes available after its start offset. \\
Bound lengths by storage capacity & Reject oversized lengths or cap them at the relevant fixed capacity before use. \\
Check multiplication before allocation & Check for overflow in element count times element size before allocating or initializing memory. \\
Check lengths before reading fields & Require enough input bytes for a fixed structure or encoded field before interpreting it. \\
Release resources on later failure & Free earlier allocations and partially constructed objects when a subsequent operation fails. \\
\bottomrule
\end{tabular}
\end{table}

\begin{figure}[t]
\centering
\begin{minipage}[t]{\columnwidth}
\footnotesize\noindent\textbf{(a) Initialize output before export: Linux cxgb3.}
\scriptsize
\begin{verbatim}
 struct ch_reg edata;

+memset(&edata, 0, sizeof(struct ch_reg));
+
 edata.cmd = CHELSIO_GET_QSET_NUM;
 edata.val = pi->nqsets;
 if (copy_to_user(useraddr, &edata, sizeof(edata)))
\end{verbatim}
\end{minipage}

\vspace{6pt}
\begin{minipage}[t]{\columnwidth}
\footnotesize\noindent\textbf{(b) Bound recursive decoding: Wireshark.}
\scriptsize
\begin{verbatim}
+depth++;
+if (depth > MP4_BOX_MAX_REC_LVL) {
 ...
+return -1;
+}
 ...
-ret = dissect_mp4_box(box_type,
-    tvb, offset, pinfo, box_tree);
+ret = dissect_mp4_box(box_type, depth,
+tvb, offset, pinfo, box_tree);
\end{verbatim}
\end{minipage}

\vspace{6pt}
\begin{minipage}[t]{\columnwidth}
\footnotesize\noindent\textbf{(c) Reserve a string terminator: ssdp-responder.}
\scriptsize
\begin{verbatim}
-char buf[MAX_PKT_SIZE];
+char buf[MAX_PKT_SIZE + 1];

 memset(buf, 0, sizeof(buf));
-len = recvfrom(sd, buf, sizeof(buf),
-    MSG_DONTWAIT, &sa, &salen);
+len = recvfrom(sd, buf, sizeof(buf) - 1,
+    MSG_DONTWAIT, &sa, &salen);
 if (len > 0) {
-buf[len] = 0;
-
\end{verbatim}
\end{minipage}
\caption{Source excerpts of three representative fix patterns. Omitted lines are marked by \texttt{...}.}
\label{fig:Pt-cluster-cases}
\end{figure}

\noindent\textbf{Repair mechanisms reflected in the clusters.}
One Jina cluster groups distinct resource-lifecycle repairs: TigerVNC deletes an owned child object, GPAC clears a freed buffer's recorded size, and a Linux failure loop combines object destruction with pointer clearing.
The cluster captures related protection concerns without implying one shared pattern.

Concrete patterns also cross broad categories: ROS, GPAC and ssdp-responder preserve string termination through bounding, initialization and capacity provision, respectively.
Conversely, a FreeRDP capacity-provision patch also performs failure cleanup.
Broad categories summarize a patch's main approach, while concrete patterns expose shared protections and cooperating edits.

\section{Part~II: Security Impact Taxonomy}
\label{sec:impact-axis}

Our security-impact study derives \BgImpactClasses{} classes from prior research and specifications, independently of patch clusters and repair categories, and labels CVE descriptions for the association analysis in Section~\ref{sec:pattern-impact}.

\subsection{Classification Unit and Evidence}
\label{sec:Ix-object}

Our security-impact study reuses the \ChRecords{}-patch corpus from our fix-pattern study (Section~\ref{sec:Ov-data}).
We label each distinct linked CVE description once and assign each patch the union of its CVEs' classes.
The association analysis retains patches with both repair-category and impact labels.

\noindent\textbf{Evidence separation.}\label{sec:Ov-separation}
Repair categories use code changes and context, with auxiliary repair explanations (Section~\ref{sec:Pt-quality}). Impact labeling instead uses cleaned CVE descriptions without patches, commit messages or CWE labels.
Remedy, provenance and severity statements are removed, and neither procedure receives the other's labels (Appendix~\ref{app:prompts}).

Each stated violation is recorded as a \emph{rule}: an operation, its object, the violated requirement, and a supporting quote (Fig.~\ref{fig:running-example}).
Distinct violations, including successive violations in a chain, receive separate classes. Access direction and resource kind remain attributes.
Thus an out-of-bounds access in a multithreaded server is memory-bounds. An additional concurrency label requires evidence of a coordination failure.
Because CVE descriptions can omit details~\cite{guo2022missing}, we distinguish insufficient evidence from stated violations outside the taxonomy.

\subsection{Taxonomy Construction}
\label{sec:Ix-design}

\begin{table*}[t]
\centering
\footnotesize
\caption{Security impact classes and their prevalence across \ChRecords{} patches.}
\label{tab:Ap-codebook}
\setlength{\tabcolsep}{4pt}
\renewcommand{\arraystretch}{1.12}
\begin{tabular*}{\textwidth}{@{\extracolsep{\fill}}>{\raggedright\arraybackslash}p{.16\textwidth} >{\raggedright\arraybackslash}p{.35\textwidth} >{\raggedright\arraybackslash}p{.33\textwidth} r@{}}
\toprule
Impact class & Security requirement & Illustrative violation & Patches, $n$ (\%) \\
\midrule
memory-bounds & Memory operations respect object bounds, alignment and overlap rules. & Copying 100 bytes into a 64-byte buffer. & 975 (25.6) \\
object-state & Use and release respect the object's current state. & Using a buffer after it has been freed. & 564 (14.8) \\
reference-and-type & References have valid targets and are used with compatible types. & Dereferencing a null pointer returned by a failed lookup. & 387 (10.1) \\
numeric & Arithmetic and conversions satisfy range, precision and domain requirements. & An allocation-size multiplication wraps around. & 291 (7.6) \\
resource-holding & Resources are released when due and held within capacity limits. & An error return leaves a file descriptor open. & 235 (6.2) \\
progress & Work progresses, stays within budget and terminates as required. & A parser loops without advancing its input position. & 160 (4.2) \\
concurrency & Threads, callbacks and signal handlers coordinate as required. & Unsynchronized counter updates lose an increment. & 130 (3.4) \\
input-acceptance & Accepted inputs and steps follow the format or protocol, with consistent interpretation. & A protocol accepts a message that skips a required handshake step. & 34 (0.9) \\
untrusted-resolution & External input is interpreted only as permitted data and resolves to allowed targets. & A supplied path escapes its allowed directory via \texttt{../}. & 67 (1.8) \\
trust-decision & Trust and access decisions satisfy identity, proof, permission and freshness requirements. & A certificate is accepted for the wrong host name. & 152 (4.0) \\
crypto-material & Cryptographic constructions, parameters and material satisfy their usage requirements. & GCM encryption reuses a nonce with the same key. & 14 (0.4) \\
sensitive-data & Sensitive data remains protected in outputs, observable behavior, storage and disposal. & A released buffer retains a private key in reusable memory. & 44 (1.2) \\
\bottomrule
\end{tabular*}
\end{table*}

\noindent\textbf{Deriving the taxonomy.}
We define security impacts by violated security requirements, distinguishing them from code defects and attack consequences~\cite{wu2020sid}.
Prior classifications of security-rule violations inform our memory, object-state, reference and permission requirements~\cite{wu2020sid}.
Other sources supply coordination~\cite{lu2008concurrency}, work-budget~\cite{crosby2003complexity}, input-recognition and interpretation~\cite{sassaman2013langsec}, protocol-transition~\cite{beurdouche2015messy}, certificate-validation~\cite{georgiev2012dangerous}, nonce-uniqueness~\cite{mcgrew2008aead}, and side-channel requirements~\cite{yarom2016cachebleed}.
CWE definitions contribute requirements such as restricted resolution and sensitive-data erasure~\cite{mitre-cwe}.

We specify each requirement's operation, object, condition and boundary examples, merge synonyms, and group \BgImpactSubtypes{} requirements into \BgImpactClasses{} classes (Table~\ref{tab:Ap-codebook}).
Grouping reflects description specificity and expected sample sizes: concurrency combines atomicity, ordering and synchronization because race reports may not distinguish them.
The taxonomy is fixed before association analysis.

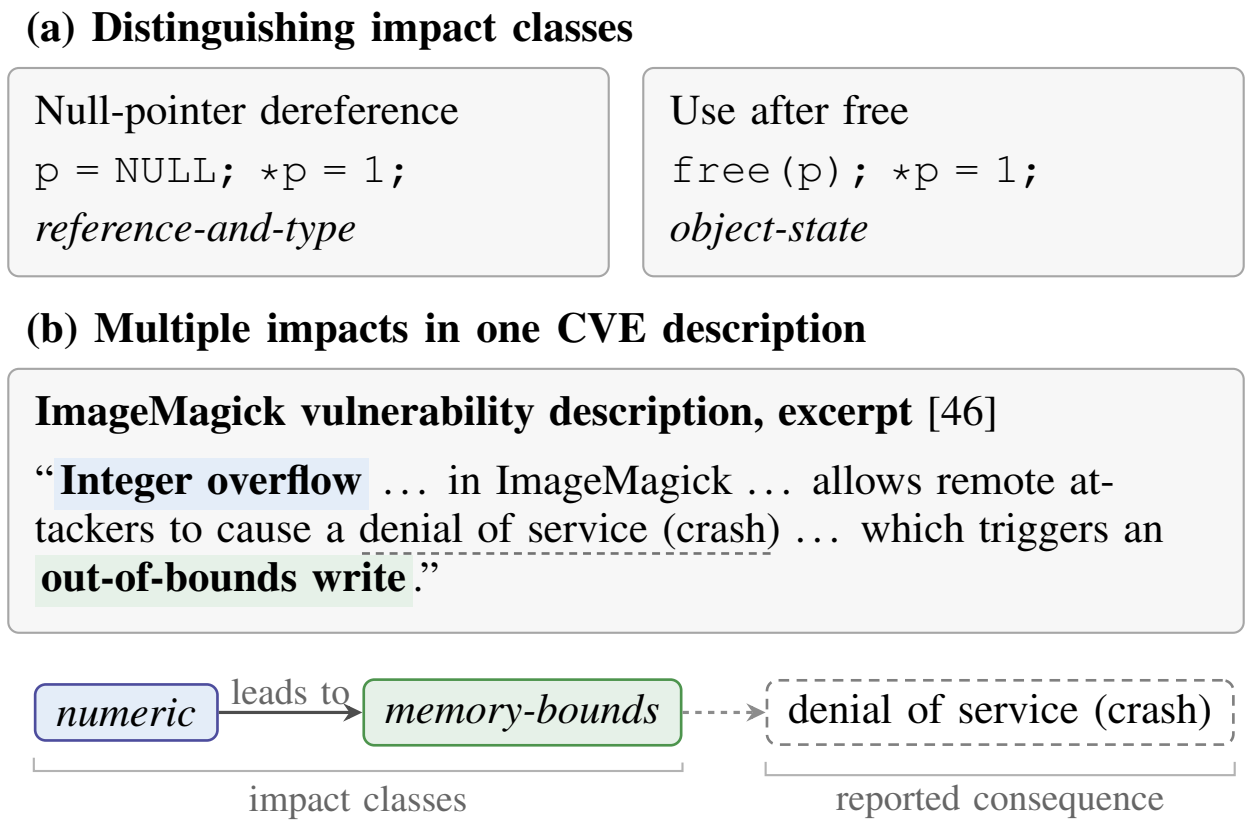
\begin{figure}[t]
\centering
\definecolor{impactnumeric}{HTML}{E4EDF8}
\definecolor{impactbounds}{HTML}{E6F1E8}
\newcommand{\IxEvidence}[2]{{\setlength{\fboxsep}{1.2pt}\colorbox{#1}{\vphantom{Ag}\textbf{#2}}}}
\newcommand{\IxConsequenceMark}[1]{\setbox0=\hbox{#1}\leavevmode\rlap{\raisebox{-2.4pt}{\color{black!55}%
  \hbox to \wd0{\leaders\hbox to 2.6pt{\hss\rule{1.5pt}{0.5pt}\hss}\hfill}}}\box0\relax}
\begin{tikzpicture}[
  x=1cm,y=1cm,font=\footnotesize,
  impactexample/.style={draw=black!35,fill=black!3,rounded corners=2pt,
    text width=3.62cm,align=left,inner sep=5pt,anchor=north west},
  chip/.style={rounded corners=2pt,inner xsep=4pt,inner ysep=2.6pt,line width=0.5pt,
    font=\footnotesize\itshape,anchor=west},
  outcome/.style={rounded corners=2pt,inner xsep=4pt,inner ysep=2.6pt,line width=0.5pt,
    draw=black!50,dash pattern=on 2pt off 1.4pt,fill=white,anchor=west},
  chain/.style={-{Stealth[length=3.6pt,width=3.4pt]},line width=0.6pt,draw=black!70},
  weak/.style={-{Stealth[length=3.6pt,width=3.4pt]},line width=0.6pt,draw=black!45,
    dash pattern=on 1.2pt off 1.2pt},
  span/.style={draw=black!30,line width=0.4pt},
  spanlabel/.style={font=\scriptsize,text=black!60,inner sep=1.5pt,anchor=north},
]
\node[anchor=west,font=\footnotesize\bfseries] at (0,0)
  {(a) Distinguishing impact classes};
\node[impactexample] at (0,-.24) {
  Null-pointer dereference\\[2pt]
  \texttt{p = NULL; *p = 1;}\\[2pt]
  \emph{reference-and-type}
};
\node[impactexample] at (4.2,-.24) {
  Use after free\\[2pt]
  \texttt{free(p); *p = 1;}\\[2pt]
  \emph{object-state}
};
\node[anchor=west,font=\footnotesize\bfseries] at (0,-1.99)
  {(b) Multiple impacts in one CVE description};
\node[impactexample,text width=7.82cm] (cvedescription) at (0,-2.23) {
  \noindent\textbf{ImageMagick vulnerability description, excerpt}~\cite{cveprogram}\\[4pt]
  ``\IxEvidence{impactnumeric}{Integer overflow} \ldots{} in ImageMagick
  \ldots{} allows remote attackers to cause a \IxConsequenceMark{denial of service (crash)}
  \ldots{} which triggers an \IxEvidence{impactbounds}{out-of-bounds write}.''
};
\node[chip,fill=impactnumeric,draw=blue!45!black!70] (cnum)
  at ([xshift=5pt,yshift=-5.2mm]cvedescription.south west) {numeric};
\node[chip,fill=impactbounds,draw=green!40!black!70] (cmem)
  at ([xshift=9.5mm]cnum.east) {memory-bounds};
\node[outcome] (ccons) at ([xshift=5.5mm]cmem.east) {denial of service (crash)};
\draw[chain] (cnum.east) -- node[above=0.2pt,font=\scriptsize,text=black!60,inner sep=1pt] {leads to} (cmem.west);
\draw[weak] (cmem.east) -- (ccons.west);
\draw[span] ([yshift=-3pt]cnum.south west) -- ++(0,-2.5pt) -| ([yshift=-3pt]cmem.south east);
\draw[span] ([yshift=-3pt]ccons.south west) -- ++(0,-2.5pt) -| ([yshift=-3pt]ccons.south east);
\node[fit=(cnum)(cmem),inner sep=0pt] (cpair) {};
\node[spanlabel] at ([yshift=-5.5pt]cpair.south) {impact classes};
\node[spanlabel] at ([yshift=-5.5pt]ccons.south) {reported consequence};
\end{tikzpicture}
\caption{Impact-class boundaries and multiple labels.}
\label{fig:Ix-classification-examples}
\end{figure}

\noindent\textbf{Boundaries and multiple labels.}
Null dereference violates reference-and-type, whereas use after free violates object-state (Fig.~\ref{fig:Ix-classification-examples}a).
Following the NIST Bugs Framework's chain perspective~\cite{bojanova2024bf}, distinct violations retain separate evidence-backed labels, as with numeric overflow leading to memory-bounds in Fig.~\ref{fig:Ix-classification-examples}b.
Alternative descriptions of one event do not create additional violations.

\subsection{Labeling and Validation}
\label{sec:Ix-labeling}

\noindent\textbf{Procedure.}
First, we extract the violated requirements, their supporting text, and any causal links stated between successive violations.
Second, we compare each requirement with the fixed class definitions and assign a label only if it matches a single class.
We use GPT-6 Luna to apply this two-stage protocol to each cleaned CVE description. We count a directed link when two causally linked violations map to different classes.

\numberedcase{case:imp-multi} The ImageMagick description (Fig.~\ref{fig:Ix-classification-examples}b) receives two labels: \emph{numeric} for the integer overflow and \emph{memory-bounds} for the resulting out-of-bounds write. Denial of service through a crash is the reported consequence.

\noindent\textbf{Labeling outcomes.}
Of \ChIxValidCVEs{} distinct usable CVE descriptions, \IxNoClassCVEs{} (\IxNoClassShareCVE{}\%) receive no impact class. Most leave the violated requirement unspecified: a report of a crash alone, for example, does not explain what went wrong. Other violations fall outside the taxonomy or cannot be mapped unambiguously.
We then assign each patch the classes found in its linked descriptions. This yields at least one class for \ChUnifiedImpactPatches{} of \ChRecords{} patches (\ChIxCoverageShare{}\%).

\noindent\textbf{Label validation.}
The authors audit \IxHumanN{} randomly sampled descriptions, establishing a reference against which automated labeling achieves 94.0\% micro-precision, 86.7\% micro-recall and \IxHumanMicroFone{} micro-F1 (Appendix~\ref{app:Ix-label-details}).

\subsection{Distribution of Security Impacts and Violation Chains}
\label{sec:Ix-scope}

Table~\ref{tab:Ap-codebook} counts each patch once per assigned class. Among all \ChRecords{} patches, memory-bounds is the most frequent class (\ChIxMemoryN{}, \ChIxMemoryShare{}\%), followed by object-state (\ChIxStateN{}, \ChIxStateShare{}\%) and reference-and-type (\ChIxReferenceN{}, \ChIxReferenceShare{}\%).

\noindent\textbf{Finding 2: When a description reports several impact classes, one violation often leads to another.}
Of the \IxChainMultiN{} CVE descriptions assigned more than one class, \IxChainLinkedN{} (\IxChainLinkedShare{}\%) contain at least one directed link under the rule in Section~\ref{sec:Ix-labeling}.
The most common link is from a numeric violation to a memory-bounds violation, as when an integer overflow leads to an out-of-bounds write (Fig.~\ref{fig:Ix-classification-examples}b). The next most common is from concurrency to object-state, as when a race leads to use after free (Appendix~\ref{app:Ix-construction}).

The ImageMagick example also illustrates what is lost when vulnerabilities are grouped by consequence alone: denial of service identifies neither the overflow nor the out-of-bounds write.
Across the corpus, descriptions mentioning denial of service collectively span all \IxDosClasses{} impact classes. Descriptions whose sole CWE label is CWE-20 (improper input validation) collectively span \IxCweTwentyClasses{}.
Identifying the specific requirement violated gives us a basis for asking which code changes restore it (Section~\ref{sec:pattern-impact}).

\section{Part~II: Associating Fix Patterns with Security Impacts}
\label{sec:pattern-impact}

We analyze thousands of security patches to identify which repair categories occur more often for a given security impact than in patches without that impact.
These associations point to candidate fix patterns for a violated security requirement. We then examine the source code to explain why patches addressing the same impact can call for different repairs.

\subsection{Association Analysis}
\label{sec:Pm-method}
\label{sec:Pm-detection}

The analysis uses \ChAssocPairedPatches{}/\ChRecords{} patches with both a primary repair category and at least one independently assigned impact class (Sections~\ref{sec:Pt-quality} and~\ref{sec:Ix-labeling}).
We retain all linked-CVE impact classes and further distinguish \emph{Check and reject} by the protected condition within the twelve repair categories.

For category $p$ and impact $i$, we compare repair prevalence within and outside the impact class:
\[
\Delta_{p,i}=P(p\mid i)-P(p\mid\neg i).
\]
Each patch counts once per comparison. Support and repository coverage admit \ChFiveTestable{} pairs.
Repository-block bootstrap intervals~\cite{efron1993bootstrap}, block-permutation tests and false-discovery correction~\cite{benjamini2001fdr} identify positive \emph{main links} meeting the interval and stratification checks in Appendix~\ref{app:assoc}.

\subsection{Repair Choices Behind the Associations}
\label{sec:Pm-assoc}
\label{sec:Pm-cases}

\begin{figure}[t]
  \centering
  \includegraphics[width=\columnwidth]{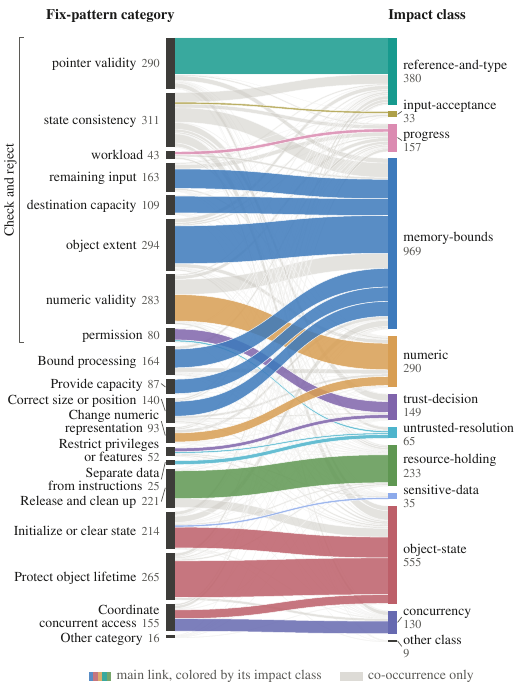}
  \caption{Repair categories and security impacts. Ribbon widths show co-occurrence counts, and colored ribbons mark main associations.}
  \label{fig:Pm-map}
\end{figure}

\noindent\textbf{Finding 3: Security impacts guide repair choices, while code context determines the applicable pattern.}
Figure~\ref{fig:Pm-map} shows \ChFiveMain{} main links, all also passing global false-discovery correction, alongside other co-occurrences.
Pointer-validity checks occur in \ChFivePointerPc{}\% of reference-and-type patches, compared with \ChFivePointerOther{}\% of patches without that class.
\emph{Release and clean up} occurs in \ChFiveCleanupPc{}\% of resource-holding patches, compared with \ChFiveCleanupOther{}\% of other patches.

Grouping patches by their code changes also brings together patches that share a security impact across project lineages.
Among same-cluster pairs from different lineages, 41.4\% shared at least one impact class in Jina, compared with 28.3\% on average when we shuffled impact labels within each repository.

Memory-bounds illustrates why context determines the repair (Fig.~\ref{fig:Pm-bounds}).
Linux \texttt{sock\_diag} rejects unsupported family numbers, whereas tpm2-tss enlarges its table from 255 to 256 entries for the supported layer \texttt{0xFF}.
Linux ib700 instead changes \texttt{>} to \texttt{>=} so its table search matches the valid 30-second maximum rather than producing index $-1$.
These repairs reject invalid requests, accommodate valid ones, or correct their lookup.

\begin{figure}[t]
\centering
\begin{minipage}[t]{\columnwidth}
\footnotesize\noindent\textbf{(a) Reject an unsupported index: Linux sock\_diag.}
\scriptsize
\begin{verbatim}
+if (req->sdiag_family >= AF_MAX)
+    return -EINVAL;
 hndl = sock_diag_lock_handler(req->sdiag_family);
\end{verbatim}
\end{minipage}

\vspace{3pt}
\begin{minipage}[t]{\columnwidth}
\footnotesize\noindent\textbf{(b) Accommodate a supported index: tpm2-tss.}
\scriptsize
\begin{verbatim}
-} layer_handler[TPM2_ERROR_TSS2_RC_LAYER_COUNT] = {
+} layer_handler[TPM2_ERROR_TSS2_RC_LAYER_COUNT + 1] = {
\end{verbatim}
\end{minipage}

\vspace{3pt}
\begin{minipage}[t]{\columnwidth}
\footnotesize\noindent\textbf{(c) Correct the index search: Linux ib700 watchdog.}
\scriptsize
\begin{verbatim}
 for (i = 0x0F; i > -1; i--)
-    if (wd_times[i] > t)
+    if (wd_times[i] >= t)
         break;
 wd_margin = i;
\end{verbatim}
\end{minipage}
\caption{Three memory-bounds repairs: reject the request, enlarge the table, or correct the lookup.}
\label{fig:Pm-bounds}
\end{figure}

The watchdog's CR cluster contains 11 patches from seven repositories, and all ten labeled members are memory-bounds.
The same \texttt{>} to \texttt{>=} edit rejects a one-past-end index in FreeRDP and radare2 but selects a valid boundary entry in the watchdog (Table~\ref{tab:Pm-boundary-context}).
Identical edits therefore implement different repair roles.

\begin{table}[t]
\centering
\footnotesize
\setlength{\tabcolsep}{3pt}
\caption{Related boundary edits with different repair roles. Conditions are abbreviated.}
\label{tab:Pm-boundary-context}
\begin{tabular}{@{}>{\raggedright\arraybackslash}p{0.48\columnwidth}>{\raggedright\arraybackslash}p{0.48\columnwidth}@{}}
\toprule
Repaired condition & Action and meaning at equality \\
\midrule
FreeRDP: \texttt{index >= number} & Return error: reject an index past the last entry. \\
radare2: \texttt{sym\_num >= nsymtab} & Skip entry: exclude an index past the last element. \\
ib700: \texttt{wd\_times[i] >= t} & Stop search: select the entry matching the requested timeout. \\
\bottomrule
\end{tabular}
\end{table}

Two Linux patches in CR cluster 52 belong to the same \emph{Release and clean up} category and make opposite edits (Fig.~\ref{fig:Pm-exit-cleanup}).
In \texttt{mwifiex\_tm\_cmd}, three error exits leave an allocated buffer unreleased (resource-holding), so the patch frees it before each return.
In \texttt{path\_openat}, \texttt{do\_tmpfile()} has already performed cleanup, so the patch skips a second release implicated in the reported use-after-free (object-state).

\begin{figure}[ht]
\centering
\begin{minipage}[t]{\columnwidth}
\footnotesize\noindent\textbf{(a) Add a missing release in \texttt{mwifiex\_tm\_cmd} (resource-holding).}
\scriptsize
\begin{verbatim}
 hostcmd = kzalloc(sizeof(*hostcmd), GFP_KERNEL);
 ...
 if (mwifiex_send_cmd(priv, 0, 0, 0, hostcmd, true)) {
     ...
+    kfree(hostcmd);
     return -EFAULT;
 }
\end{verbatim}
\end{minipage}

\vspace{3pt}
\begin{minipage}[t]{\columnwidth}
\footnotesize\noindent\textbf{(b) Skip repeated cleanup in \texttt{path\_openat} (object-state).}
\scriptsize
\begin{verbatim}
 error = do_tmpfile(dfd, pathname, nd, flags, op,
                    file, &opened);
-goto out;
+goto out2;
 ...
 out:
 path_cleanup(nd);
+out2:
\end{verbatim}
\end{minipage}
\caption{Two patches in CR cluster 52 add a missing release (a) or skip a repeated one (b).}
\label{fig:Pm-exit-cleanup}
\end{figure}

For recursion limits, the operation being guarded explains which impact the pattern addresses.
open62541 bounds recursion depth before indexing \texttt{ctx->commaNeeded}, protecting memory-bounds, while Linux Netlink bounds depth to stop indefinite recursion through policy back references, protecting progress.
Impact narrows repair choices, clusters locate relevant examples, and source context determines applicability.

\section{Part~III: Retrieval-Augmented Repair}
\label{sec:repair-check}

We study how the choice of fix examples affects an LLM's ability to identify the unsafe operation and repair the vulnerability.
For retrieval-augmented generation, we use patch clusters and impact--pattern associations to retrieve relevant examples from a repair knowledge base.
We compare example-selection methods, examine the resulting plans and repairs, and test the roles of query quality, project lineage and code representation.

\begin{figure}[t]
\centering
\includegraphics[width=\columnwidth]{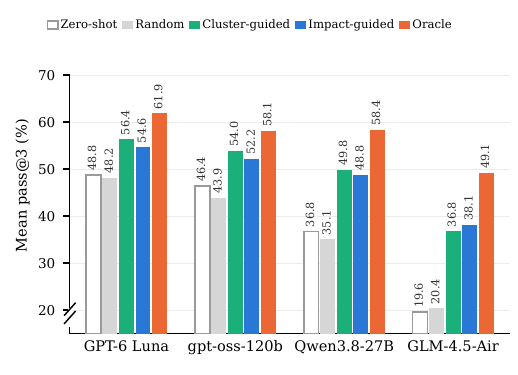}
\caption{Pass@3 across \ChSixTargets{} targets of mixed difficulty. The vertical axis starts at 15\%.}
\label{fig:repair-bars}
\end{figure}

\subsection{Experimental Design}
\label{sec:Rc-setup}

\noindent\textbf{Targets and examples.}
Our vulnerability-repair study evaluates a separate held-out set of \ChSixTargets{} C/C++ repair targets from \ChSixRepos{} repositories in \ChSixSourceGroups{} project lineages.
The test patches and all other members of their repair families were excluded from clustering.
We stratify the targets by developer-patch complexity into \ChSixEasy{} easy, \ChSixMedium{} medium and \ChSixHard{} hard targets: easy fixes make a short, localized change, while hard fixes involve multiple functions, larger or scattered edits, or several newly introduced identifiers.
Appendix~\ref{app:repair} details benchmark construction and the stratification rules.
For retrieval, we reuse the patch clusters from our fix-pattern study and the associations from our security-impact study, with an expanded knowledge-base pool of \ChSixPoolPatches{} security patches.
Patches beyond the \ChRecords{} clustering cohort are assigned to its existing clusters rather than reclustering the pool.
We draw examples from other repair families, including those in the target's own project lineage.

\noindent\textbf{Retrieval methods and comparison conditions.}
In these experiments, a fix example contains the pre-fix code and diff of one function.
Each few-shot prompt includes three fix examples.
\emph{Zero-shot} supplies no examples, whereas \emph{Random} samples eligible repair families.
\emph{Cluster-guided} selects examples by similarity to a query fix. It searches the corresponding patch cluster first and ranks examples by similarity between code changes.
\emph{Impact-guided} uses the impact--pattern associations to find the three repair categories most common for the target's security impact in the repair knowledge base. It retrieves fix examples from these categories, looking first in the query fix's cluster and ranking examples by code-change similarity.
Both methods use CR clusters and the developer fix as the query in the main comparison. Section~\ref{sec:Rc-ablation} also evaluates model-generated queries.
\emph{Oracle} selects carefully screened fix examples from a broad candidate set.
It prioritizes examples implementing the target's repair mechanism and fills any remaining slots with relevant fix examples that share a necessary repair element.

\noindent\textbf{Generation and evaluation.}
We evaluate four LLMs. GPT-6 Luna~\cite{openai2026gpt6luna} is a commercial frontier model, and gpt-oss-120b~\cite{openai2025gptoss}, Qwen3.8-27B and GLM-4.5-Air~\cite{glm2025glm45} are open-weight models of different sizes. Each model first produces a repair plan, then generates corrected functions in the same conversation.
The repair prompt supplies the vulnerable functions and selected examples, without the target vulnerability description or location hints from the developer fix.
Our evaluation protocol accepts a candidate if it blocks the same dangerous path as the developer fix and produces an equivalent effect on control flow. GPT-6 Luna applies this criterion to the source changes with the generator and retrieval method hidden.
We generate three candidates per example group and count the group as successful if any candidate passes (pass@3).

\subsection{Repair Effectiveness}
\label{sec:Rc-results}
\label{sec:Rc-plans}

\begin{table}[t]
\centering
\footnotesize
\caption{Pass@3 (\%) by edit-complexity stratum, with the difference to Random in the same stratum ($\Delta$R, percentage points). Bold marks the highest rate per generator and stratum, and cell shading follows pass@3.}
\label{tab:repair-difficulty}
\definecolor{ramp1}{HTML}{FFFFFF}\definecolor{ramp2}{HTML}{EEF4FB}\definecolor{ramp3}{HTML}{D9E6F6}\definecolor{ramp4}{HTML}{BDD3EE}\definecolor{ramp5}{HTML}{9EBEE6}
\definecolor{dpos}{HTML}{1A7F37}\definecolor{dneg}{HTML}{B3261E}\definecolor{dgrey}{HTML}{7A7A7A}
\setlength{\tabcolsep}{1.6pt}\renewcommand{\arraystretch}{1.04}
\begin{tabular}{@{}ll r@{\,}l r@{\,}l r@{\,}l@{}}
\toprule
Generator & Method & \multicolumn{2}{c}{Easy (111)} & \multicolumn{2}{c}{Medium (107)} & \multicolumn{2}{c}{Hard (73)} \\
 &  & \multicolumn{1}{c}{\scriptsize pass@3} & \multicolumn{1}{c}{\scriptsize $\Delta$R} & \multicolumn{1}{c}{\scriptsize pass@3} & \multicolumn{1}{c}{\scriptsize $\Delta$R} & \multicolumn{1}{c}{\scriptsize pass@3} & \multicolumn{1}{c}{\scriptsize $\Delta$R} \\
\midrule
\multirow{5}{*}{GPT-6 Luna} & Zero-shot & \cellcolor{ramp5}65.8 & {\color{dpos}\scriptsize $+2.4$} & \cellcolor{ramp3}39.3 & {\color{dneg}\scriptsize $-2.8$} & \cellcolor{ramp3}37.0 & {\color{dpos}\scriptsize $+2.7$} \\
 & Random & \cellcolor{ramp4}63.4 & -- & \cellcolor{ramp3}42.1 & -- & \cellcolor{ramp2}34.2 & -- \\
 & Cluster-guided & \cellcolor{ramp5}73.9 & {\color{dpos}\scriptsize $+10.5$} & \cellcolor{ramp4}53.3 & {\color{dpos}\scriptsize $+11.2$} & \cellcolor{ramp2}34.2 & {\color{dgrey}\scriptsize $0.0$} \\
 & Impact-guided & \cellcolor{ramp5}72.1 & {\color{dpos}\scriptsize $+8.7$} & \cellcolor{ramp3}48.6 & {\color{dpos}\scriptsize $+6.5$} & \cellcolor{ramp3}37.0 & {\color{dpos}\scriptsize $+2.7$} \\
 & Oracle & \cellcolor{ramp5}\textbf{76.6} & {\color{dpos}\scriptsize $+13.2$} & \cellcolor{ramp4}\textbf{57.9} & {\color{dpos}\scriptsize $+15.9$} & \cellcolor{ramp3}\textbf{45.2} & {\color{dpos}\scriptsize $+11.0$} \\
\midrule
\multirow{5}{*}{gpt-oss-120b} & Zero-shot & \cellcolor{ramp4}57.7 & {\color{dgrey}\scriptsize $0.0$} & \cellcolor{ramp3}43.9 & {\color{dpos}\scriptsize $+4.7$} & \cellcolor{ramp2}32.9 & {\color{dpos}\scriptsize $+3.2$} \\
 & Random & \cellcolor{ramp4}57.7 & -- & \cellcolor{ramp3}39.3 & -- & \cellcolor{ramp2}29.7 & -- \\
 & Cluster-guided & \cellcolor{ramp5}\textbf{75.7} & {\color{dpos}\scriptsize $+18.0$} & \cellcolor{ramp3}47.7 & {\color{dpos}\scriptsize $+8.4$} & \cellcolor{ramp2}30.1 & {\color{dgrey}\scriptsize $+0.5$} \\
 & Impact-guided & \cellcolor{ramp5}69.4 & {\color{dpos}\scriptsize $+11.7$} & \cellcolor{ramp3}47.7 & {\color{dpos}\scriptsize $+8.4$} & \cellcolor{ramp2}32.9 & {\color{dpos}\scriptsize $+3.2$} \\
 & Oracle & \cellcolor{ramp5}72.1 & {\color{dpos}\scriptsize $+14.4$} & \cellcolor{ramp4}\textbf{59.8} & {\color{dpos}\scriptsize $+20.6$} & \cellcolor{ramp2}\textbf{34.2} & {\color{dpos}\scriptsize $+4.6$} \\
\midrule
\multirow{5}{*}{Qwen3.8-27B} & Zero-shot & \cellcolor{ramp3}49.5 & {\color{dgrey}\scriptsize $0.0$} & \cellcolor{ramp2}34.6 & {\color{dpos}\scriptsize $+6.9$} & \cellcolor{ramp2}20.5 & {\color{dneg}\scriptsize $-3.2$} \\
 & Random & \cellcolor{ramp3}49.5 & -- & \cellcolor{ramp2}27.7 & -- & \cellcolor{ramp2}23.7 & -- \\
 & Cluster-guided & \cellcolor{ramp5}\textbf{74.8} & {\color{dpos}\scriptsize $+25.2$} & \cellcolor{ramp3}38.3 & {\color{dpos}\scriptsize $+10.6$} & \cellcolor{ramp2}28.8 & {\color{dpos}\scriptsize $+5.0$} \\
 & Impact-guided & \cellcolor{ramp5}73.0 & {\color{dpos}\scriptsize $+23.4$} & \cellcolor{ramp3}38.3 & {\color{dpos}\scriptsize $+10.6$} & \cellcolor{ramp2}27.4 & {\color{dpos}\scriptsize $+3.7$} \\
 & Oracle & \cellcolor{ramp5}\textbf{74.8} & {\color{dpos}\scriptsize $+25.2$} & \cellcolor{ramp4}\textbf{56.1} & {\color{dpos}\scriptsize $+28.3$} & \cellcolor{ramp3}\textbf{37.0} & {\color{dpos}\scriptsize $+13.2$} \\
\midrule
\multirow{5}{*}{GLM-4.5-Air} & Zero-shot & \cellcolor{ramp2}27.9 & {\color{dpos}\scriptsize $+1.8$} & \cellcolor{ramp1}17.8 & {\color{dgrey}\scriptsize $+0.6$} & \cellcolor{ramp1}9.6 & {\color{dneg}\scriptsize $-6.8$} \\
 & Random & \cellcolor{ramp2}26.1 & -- & \cellcolor{ramp1}17.1 & -- & \cellcolor{ramp1}16.4 & -- \\
 & Cluster-guided & \cellcolor{ramp4}52.3 & {\color{dpos}\scriptsize $+26.1$} & \cellcolor{ramp2}28.0 & {\color{dpos}\scriptsize $+10.9$} & \cellcolor{ramp2}26.0 & {\color{dpos}\scriptsize $+9.6$} \\
 & Impact-guided & \cellcolor{ramp3}47.7 & {\color{dpos}\scriptsize $+21.6$} & \cellcolor{ramp3}36.4 & {\color{dpos}\scriptsize $+19.3$} & \cellcolor{ramp2}26.0 & {\color{dpos}\scriptsize $+9.6$} \\
 & Oracle & \cellcolor{ramp4}\textbf{63.1} & {\color{dpos}\scriptsize $+36.9$} & \cellcolor{ramp3}\textbf{48.6} & {\color{dpos}\scriptsize $+31.5$} & \cellcolor{ramp2}\textbf{28.8} & {\color{dpos}\scriptsize $+12.3$} \\
\midrule
\multirow{5}{*}{\emph{Mean of four}} & Zero-shot & \cellcolor{ramp4}50.2 & {\color{dpos}\scriptsize $+1.1$} & \cellcolor{ramp2}33.9 & {\color{dpos}\scriptsize $+2.3$} & \cellcolor{ramp2}25.0 & {\color{dneg}\scriptsize $-1.0$} \\
 & Random & \cellcolor{ramp3}49.2 & -- & \cellcolor{ramp2}31.5 & -- & \cellcolor{ramp2}26.0 & -- \\
 & Cluster-guided & \cellcolor{ramp5}69.1 & {\color{dpos}\scriptsize $+20.0$} & \cellcolor{ramp3}41.8 & {\color{dpos}\scriptsize $+10.3$} & \cellcolor{ramp2}29.8 & {\color{dpos}\scriptsize $+3.8$} \\
 & Impact-guided & \cellcolor{ramp5}65.5 & {\color{dpos}\scriptsize $+16.4$} & \cellcolor{ramp3}42.8 & {\color{dpos}\scriptsize $+11.2$} & \cellcolor{ramp2}30.8 & {\color{dpos}\scriptsize $+4.8$} \\
 & Oracle & \cellcolor{ramp5}\textbf{71.6} & {\color{dpos}\scriptsize $+22.4$} & \cellcolor{ramp4}\textbf{55.6} & {\color{dpos}\scriptsize $+24.1$} & \cellcolor{ramp3}\textbf{36.3} & {\color{dpos}\scriptsize $+10.3$} \\
\bottomrule
\end{tabular}

\end{table}

\noindent\textbf{Finding 4: Cluster-guided and Impact-guided retrieval improve repair even for a frontier model, with the largest gains for models with lower baseline repair rates.}
Both methods improve pass@3 over Random for all four generators (Fig.~\ref{fig:repair-bars}), by \ChSixMeanClusterGain{} and \ChSixMeanImpactGain{} percentage points on average, respectively.
For GPT-6 Luna, the gains are 8.1 and 6.4 points. For Qwen3.8-27B, pass@3 rises from 35.1\% with Random to 49.8\% and 48.8\%, and for GLM-4.5-Air it rises from 20.4\% to 36.8\% and 38.1\%.
These gains come from changing which three examples the generator receives, without fine-tuning it or adding a generation step.

Table~\ref{tab:repair-difficulty} shows how this benefit varies with edit complexity. Across the four models, Random achieves 49.2\% on easy targets, 31.5\% on medium targets and 26.0\% on hard targets.
At each difficulty, the four-model mean is higher for Cluster-guided, Impact-guided and Oracle than for Random.
The gains for Cluster-guided and Impact-guided fall from \ChSixEasyClusterGain{} and \ChSixEasyImpactGain{} points on easy targets to \ChSixMediumClusterGain{} and \ChSixMediumImpactGain{} on medium targets and \ChSixHardClusterGain{} and \ChSixHardImpactGain{} on hard targets. Oracle's gain peaks on medium targets (\ChSixMediumOracleGain{} points) and reaches \ChSixHardOracleGain{} points on hard targets.

\noindent\textbf{Relevant examples help some open-weight models approach frontier-model repair rates.}
On easy targets, Cluster-guided retrieval yields 74.8\% pass@3 for Qwen3.8-27B and 75.7\% for gpt-oss-120b, compared with 73.9\% for GPT-6 Luna under the same method.
On medium targets, Impact-guided retrieval brings gpt-oss-120b to 47.7\%, within 0.9 percentage points of GPT-6 Luna's 48.6\%.
These observed similarities show that relevant repair knowledge can help open-weight models close the performance gap on specific difficulty levels, although GLM remains below GPT-6 Luna and larger gaps persist on hard targets.

\textbf{Simply adding examples provides no aggregate improvement over Zero-shot, and Random yields lower observed repair rates for three of the four generators}.
Fix examples are therefore useful when their content guides the target repair.
Unrelated examples can leave the relevant defect unresolved or direct changes toward a different problem.
Our clustering and impact--pattern map support retrieval of relevant examples from the repair knowledge base: clusters organize the example pool, while impact associations prioritize repair categories relevant to the target's security requirement.
The retrieved examples illustrate how checks, initialization or resource management restore that requirement, providing implementations the generator can adapt to the target's context.
Oracle provides an optimistic reference for carefully screened examples, reaching \ChSixMeanOracle{}\%.

\begin{figure*}[!t]
\centering
\begin{minipage}[t]{0.48\textwidth}
\footnotesize\textbf{(a) Initialize the whole object before export.}
\par\smallskip
\scriptsize\textsf{Retrieved fix example: LibVNCServer}
\begin{verbatim}
 rfbServerCutTextMsg sct;
+memset((char *)&sct, 0, sizeof(sct));
 ...
 sct.type = rfbServerCutText;
 sct.length = Swap32IfLE(len);
 ...
 if (rfbWriteExact(cl, (char *)&sct,
                  sz_rfbServerCutTextMsg) < 0) {
\end{verbatim}
\textsf{Generated target fix: Linux EQL}
\begin{verbatim}
 master_config_t mc;
+memset(&mc, 0, sizeof(mc));
 ...
 mc.max_slaves = eql->max_slaves;
 mc.min_slaves = eql->min_slaves;
 if (copy_to_user(mcp, &mc,
                  sizeof(master_config_t)))
     return -EFAULT;
\end{verbatim}
\end{minipage}\hfill
\begin{minipage}[t]{0.48\textwidth}
\footnotesize\textbf{(b) Tie allocation to its cleanup scope.}
\par\smallskip
\scriptsize\textsf{Retrieved fix example: Wireshark TN3270}
\begin{verbatim}
 tn3270_info = wmem_new(wmem_file_scope(),
                        tn3270_conv_info_t);
 ...
-copy_address(...);
+copy_address_wmem(wmem_file_scope(), ...);
 ...
 conversation_add_proto_data(...);
\end{verbatim}
\textsf{Generated target fix: Wireshark PCP}
\begin{verbatim}
-pcp_conv_info = (pcp_conv_info_t*)g_malloc(...);
+pcp_conv_info = wmem_new(wmem_file_scope(),
+                         pcp_conv_info_t);
 conversation_add_proto_data(...);
\end{verbatim}
\end{minipage}
\caption{Retrieved fix examples and generated repairs: (a) GLM initializes an exported object, and (b) Qwen3.8 binds allocation to its cleanup scope. Ellipses omit code.}
\label{fig:repair-cases}
\end{figure*}

To understand these gains, we assess each attempt's repair plan separately from its generated code (Table~\ref{tab:repair-plans}).
\emph{Problem identified} checks whether a plan identifies the unsafe operation and missing protection, while \emph{Complete plan} checks whether it includes all required repair steps.
Compared with Random, Cluster-guided improves these scores from 33.4\% to 47.0\% and from 21.6\% to 31.8\%, respectively.
Impact-guided reaches 45.9\% and 31.2\%, while Oracle reaches 57.0\% and 39.3\%.
Both guided methods improve problem identification more than plan completeness, suggesting that relevant examples help models recognize what needs fixing.
Planning a complete repair may still require information about callers or project-specific behavior that the supplied functions do not provide.

\begin{table}[t]
\centering
\footnotesize
\setlength{\tabcolsep}{3pt}
\caption{Per-attempt plan quality (\%), averaged across targets and generators.}
\label{tab:repair-plans}
\renewcommand{\arraystretch}{.90}
\begin{tabular*}{.90\columnwidth}{@{}l@{\extracolsep{\fill}}rr@{}}
\toprule
Method & Problem identified & Complete plan \\
\midrule
Zero-shot & 33.2 & 22.5 \\
Random & 33.4 & 21.6 \\
Cluster-guided & 47.0 & 31.8 \\
Impact-guided & 45.9 & 31.2 \\
Oracle & \textbf{57.0} & \textbf{39.3} \\
\bottomrule
\end{tabular*}

\end{table}

We further reviewed \RepairHumanN{} candidates counted as successful and judged \RepairHumanPathShare{}\% (\RepairHumanPathN{}/\RepairHumanN{}) to block the target dangerous path (Appendix~\ref{app:repair}).
We next illustrate how retrieved fix examples inform repair plans and guide the generated patches.

\numberedcase{case:repair-initialization} A shared initialization pattern transfers across output interfaces.
In Linux EQL, \texttt{eql\_g\_master\_cfg} fills two fields of a stack object and copies the whole object to user space, exposing its uninitialized 16-byte name field.
In a recorded Random trial, GLM added a destination-pointer check but left the object uninitialized.
Cluster-guided retrieved whole-object initialization examples, including a LibVNCServer message sent over a socket (Fig.~\ref{fig:repair-cases}a).
GLM's plan explicitly connected those examples to uninitialized-data exposure and required clearing the entire object before assigning fields.
The generated \texttt{memset} implements the developer's repair, and GLM repairs the target the same way when same-lineage examples are excluded.

\numberedcase{case:repair-lifecycle} Impact associations direct example selection toward the required memory-lifecycle repair.
In Wireshark's PCP dissector, conversation state allocated with \texttt{g\_malloc} is not reclaimed through the file's memory-management scope.
With Random or Cluster-guided examples, Qwen3.8 instead adds packet-length checks and leaves this allocation unchanged.
The target's resource-holding impact favors cleanup-related categories. Impact-guided therefore retrieves a TN3270 fix from a neighboring cluster that places address-copy allocations under \texttt{wmem\_file\_scope} (Fig.~\ref{fig:repair-cases}b).
Its context also allocates conversation state in that scope.
Qwen3.8 cites this pattern in its plan and generates the corresponding scope-managed allocation, matching the developer's repair.

\subsection{Ablation Studies}
\label{sec:Rc-ablation}

\begin{figure}[!t]
\centering
\includegraphics[width=\columnwidth]{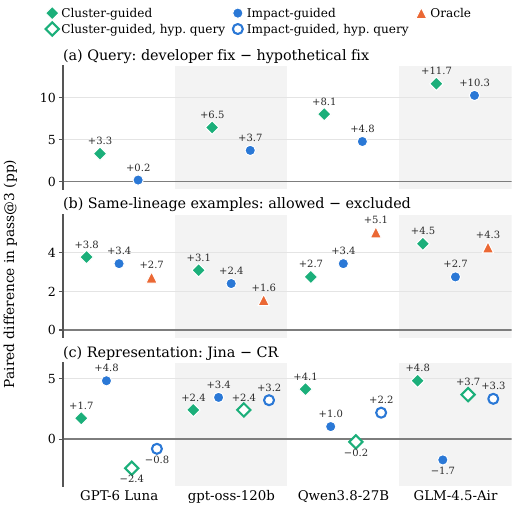}
\caption{Effects of query quality, same-lineage example availability and patch representation on pass@3. Panels use separate scales.}
\label{fig:repair-ablation}
\end{figure}

We examine how query quality, project lineage and code representation affect example-guided repair (Fig.~\ref{fig:repair-ablation}).
Appendix~\ref{app:repair} reports the paired differences across the four models.

\noindent\textbf{Sensitivity to the query fix.}
We ask each model to generate a preliminary fix and use it as the retrieval query. Both retrieval methods retain gains over Zero-shot and Random on the four-model mean.
Switching queries reduces pass@3 less for Impact-guided than for Cluster-guided in each of the four generators (Fig.~\ref{fig:repair-ablation}a).
Changing the query can change the starting cluster and the ranking of examples. For Impact-guided, the target's security impact still identifies the same candidate repair categories, even when the preliminary fix points to a different part of the patch collection.

\noindent\textbf{Effect of shared project lineage.}
We compare retrieval with and without examples from the target's project lineage, including related repositories.
Allowing these examples raises mean pass@3 by 3.5 percentage points for Cluster-guided, 3.0 for Impact-guided and 3.4 for Oracle (Fig.~\ref{fig:repair-ablation}b). Random changes by only $-0.3$ points.
When same-lineage examples are excluded, Impact-guided still exceeds Random by \ChSixXplcImpactDevGain{} points with developer queries and by \ChSixXplcImpactGain{} points with hypothetical queries (Table~\ref{tab:repair-ablation}).
Shared lineage contributes useful context, but the retained gains show that fix examples also provide guidance across lineages.

\noindent\textbf{Changing the representation.}
Replacing CR by the Jina representation and clustering pipeline of Section~\ref{sec:Pt-representations} preserves the gains over Zero-shot and Random for both guided methods on the four-model mean, under both developer and hypothetical queries (Fig.~\ref{fig:repair-ablation}c).
The impact prior also remains useful under Jina, adding \ChSixJinaFilterHyp{} points to Cluster-guided with hypothetical queries.
The consistent benefits across the two representations support organizing security patches through repair patterns and their security-impact associations as a basis for retrieving relevant fix examples from a knowledge base for retrieval-augmented generation.

\section{Discussion}
\label{sec:discussion}

Our findings show how security requirements can serve as a common basis for organizing repair knowledge and applying it to new vulnerabilities.

\noindent\textbf{Security impacts guide repair choices.}
Expressing security impact through violated requirements gives repair a concrete objective.
The impact--pattern map connects that objective to candidate repair approaches, while the target's code context determines their applicability.
Security impact thus guides a decision central to repair generation: which repair knowledge is relevant to the target.

\noindent\textbf{Fix patterns explain what clustering captures.}
Organizing this knowledge requires patch similarities that reflect how fixes work.
The patterns identified in the clusters connect different code forms through the operations they protect and the security requirements they restore.
By making shared repair mechanisms explicit, these patterns explain how patch similarity can support the search for relevant fix examples.

\noindent\textbf{Relevant examples help models diagnose the defect.}
Even when a preliminary fix leads retrieval to a different cluster, the target's security impact narrows the search to the same candidate repair categories.
This helps explain why switching from developer fixes to preliminary fixes generated with change-location hints reduces pass@3 less for Impact-guided than for Cluster-guided retrieval across all four models (Fig.~\ref{fig:repair-ablation}a).
Compared with Random, both guided methods improve identification of the unsafe operation and its missing protection more than they improve plan completeness (Table~\ref{tab:repair-plans}).
The observed repair gains illustrate the practical value of a knowledge base organized around fix patterns and security impacts for retrieval-augmented patch generation.

\noindent\textbf{Future directions.}
Representations that capture repair mechanisms more precisely could preserve the relationships between a patch, the operations it protects and its necessary context across different code forms and projects.
The gap to Oracle motivates ranking fix examples by how well they illustrate the protection missing from the target code and combining complementary examples for coordinated repairs.
Future work could test whether pairing vulnerable code with such examples during post-training helps models recognize when the same protection is needed elsewhere.
These directions converge on a broader goal: a repair workflow in which security requirements connect diagnosis, retrieval of relevant fix examples, generation and validation.
Feedback from applying and checking candidate repairs could refine the diagnosis and guide subsequent retrieval from the repair knowledge base, extending the use of impact--pattern associations into an iterative repair process.

\section{Related Work}
\label{sec:related}

\noindent\textbf{Mining fix patterns.}
Pattern mining supports the construction of repair rules and transformation templates.
FixMiner~\cite{koyuncu2020fixminer} iteratively clusters AST-level edit scripts of atomic changes and integrates the patterns into a repair prototype.
Getafix~\cite{bader2019getafix} learns hierarchical patterns for static-analyzer warnings and ranks fixes by context. Liu et al.~\cite{liu2017findbugs} cluster learned features of violations and their fixes.
Revisar~\cite{rolim2018revisar} mines recurring edits from revision histories, and SPINFER~\cite{serrano2020spinfer} infers semantic-patch rules for large-scale Linux changes.
VuRLE~\cite{ma2017vurle} learns edit and context templates from vulnerable and repaired code for vulnerability detection and repair.
Pan et al.~\cite{pan2009bugfixpatterns}, Zhong and Su~\cite{zhong2015realbugfixes}, and Sobreira et al.~\cite{sobreira2018dissection} characterize bug fixes and their recurring patterns. PPD~\cite{madeiral2018ppd} and Coming~\cite{martinez2019coming} detect predefined patterns in patches and commits.
Yang et al.~\cite{yang2022pythonfix} provide a Python fix catalog that informed our manual fix-action labels.
Our study characterizes the security-repair mechanisms reflected in patch clusters and relates their applicability to the security requirements restored.

\noindent\textbf{Patch representations and analysis units.}
TreeCen~\cite{hu2022treecen} and DSFM~\cite{feng2024dsfm} use structural features for clone detection, while Feng et al.~\cite{feng2024token} study token-based comparison.
GraphCodeBERT~\cite{guo2021graphcodebert} learns data-flow-aware code representations.
For code changes, CC2Vec~\cite{hoang2020cc2vec} models added and removed code, CCRep~\cite{liu2023ccrep} relates changed tokens to context, and GraphSPD~\cite{wang2023graphspd} combines pre- and post-patch code property graphs for security-patch detection.
CodeReviewer~\cite{li2022codereviewer} is pretrained for code-change and review tasks, and Jina~\cite{kryvosheieva2025jina} for code retrieval.
We evaluate patch representations constructed from these frozen encoders against AST-edit summaries for organizing repair mechanisms.
Li and Paxson~\cite{li2017security} document the locality of security patches, while Fixseeker~\cite{cheng2026fixseeker} examines correlations between hunks.
Herbold et al.~\cite{herbold2022tangling} study mixed changes in bug-fixing commits, and Atomizer~\cite{zhu2026atomizer} separates commit intents.
Sejfia et al.~\cite{sejfia2024mbu} examine the challenges of detecting vulnerabilities that span multiple code units.
Our analysis aggregates each patch's available changes across functions and files into one clustering point, supporting source analysis of repairs involving several locations.

\noindent\textbf{Security impact characterization.}
SID~\cite{wu2020sid} uses security-rule comparisons and patch models to identify the impacts addressed by Linux kernel patches.
DUALLM~\cite{li2026whatdotheyfix} combines commit-message and diff analysis with sliced code to classify memory bugs.
SPAIN~\cite{xu2017spain} derives vulnerability and patch patterns from binary changes for security-patch analysis.
PatchDB~\cite{wang2021patchdb} characterizes security-patch changes and illustrates repair patterns as a potential use of its dataset.
We characterize violated security requirements from vulnerability descriptions separately from code-based repair categories, then systematically measure their associations.
Clustered examples explain how code context distinguishes concrete repair choices for the same impact, connecting impact characterization with the applicability of repair knowledge.

\noindent\textbf{Knowledge-guided vulnerability repair.}
Pearce et al.~\cite{pearce2023zeroshot} evaluate LLMs for zero-shot vulnerability repair, while CEDAR~\cite{nashid2023cedar} retrieves demonstrations for code-related few-shot tasks, including program repair.
Given CWE types and vulnerable locations, APPATCH~\cite{nong2025appatch} selects examples by inferred root cause.
VulMaster~\cite{zhou2024vulmaster} incorporates CWE knowledge and examples into a trained repair model.
VulKey~\cite{li2026vulkey} organizes domain-specific repair knowledge by CWE type, syntactic actions and semantic key elements.
VulMatch~\cite{cao2025vulmatch} clusters repair representations into prototypes to guide a fine-tuned repair model.
CWEFT~\cite{carbonell2026cweft} reports benefits from naming the CWE class in Java repair, whereas VulnRepairEval~\cite{wang2025vulnrepaireval} finds limited gains from enhanced prompting under exploit-based validation.
Our study investigates security requirements as a basis for retrieving relevant fix examples from a repair knowledge base to support retrieval-augmented generation.
We connect patch clusters with measured impact--pattern associations and evaluate how the resulting fix examples help LLMs understand and repair new vulnerabilities.

\section{Conclusion}
\label{sec:conclusion}

Our large-scale study identifies recurring fix patterns across syntactically different patches and relates them to \BgImpactClasses{} security impact classes.
The resulting map connects violated requirements with candidate repair strategies, while code context determines their applicability.

Retrieving relevant examples from this knowledge base improves patch generation across four LLMs, with particularly large gains for Qwen and GLM.
Under the same retrieval methods, Qwen and gpt-oss reach repair rates similar to GPT-6 Luna on easy targets, and gpt-oss approaches Luna on medium targets.
Random examples instead reduce repair rates for three models, emphasizing that relevance matters.
The remaining gaps on harder repairs motivate further research on retrieving and applying fix-pattern and security-impact knowledge to prevent exploitation while preserving intended functionality.

\bibliographystyle{IEEEtran}
\bibliography{references}

\appendices
\makeatletter
\setlength{\@fptop}{0pt}
\setlength{\@fpsep}{12pt}
\setlength{\@fpbot}{0pt plus 1fil}
\makeatother
\raggedbottom
\section{Patch Clustering and Evaluation}
\label{app:discovery}\label{app:clustering}\label{app:design}\label{app:deviations}

\noindent\textbf{Cohort and representations.}
Records sharing a repository and fixing commit are consolidated, and equivalent backports form repair families.
The \ChRecords{}-patch cohort includes \ChPartial{} partially recovered samples.
Source groups link repositories through recorded forks, patch dependencies and identical normalized diffs. These links define the provenance groups used in the semantic evaluation.
CR and Jina use 512- and 8,192-token encoder calls, respectively. Long-input readouts are averaged across non-overlapping chunks, weighted by code-token counts.
Geometric scores use each pipeline's working space, while semantic scores use the same frozen references.

\noindent\textbf{Semantic evaluation.}
$T_3$ is patch-weighted coverage by each cluster's three largest repair categories.
$P_{\times}$ measures primary-label agreement among cross-source-group pairs using 22 condition-refined labels, with undetermined labels excluded.
For query $q$, let $G(q)$ be its cluster, $P_q$ its known same-pattern partners and $D_q$ its reviewed same-source-group distractors:
\[
C_q=\frac{|P_q\cap G(q)|}{|P_q|},\qquad
M_q=\frac{|D_q\cap G(q)|}{|D_q|}.
\]
$C$ averages over \ChFineQueries{} queries with \ChFinePairs{} directed partner pairs. $M$ averages over \ChMixQueries{} queries with \ChMixPairs{} distractor pairs.
Unjudged relationships are not negatives. Distractors reject the specified pattern rather than every possible shared pattern.

\noindent\textbf{Labeling and human checks.}
Qwen3.8 labels code-only and code-plus-repair-context views, and GPT-6 Luna adjudicates disagreements and low-confidence cases.
Table~\ref{tab:Ap-human-checks} summarizes the human checks across the study.
Each audit has one software-security annotator.
The category audit combines 30 patches selected for category coverage and boundary cases with 70 randomly selected repair families.
The impact audit randomly samples descriptions and includes empty class sets.
The repair audit samples three accepted candidates per generator--method combination.

\begin{table}[H]
\centering\footnotesize
\setlength{\tabcolsep}{3pt}
\renewcommand{\arraystretch}{1.12}
\caption{Human audit results.}
\label{tab:Ap-human-checks}
\begin{tabular}{@{}>{\raggedright\arraybackslash}p{.23\columnwidth}>{\raggedright\arraybackslash}p{.23\columnwidth}>{\raggedright\arraybackslash}p{.47\columnwidth}@{}}
\toprule
Audit & Sample & Result \\
\midrule
Repair categories & 100 patches & Primary-label agreement: 79/100 \\
Impact classes & \IxHumanN{} descriptions & Exact class-set agreement: \IxHumanAgreeShare{}\%, micro-F1: \IxHumanMicroFone{} \\
Repair assessment & \RepairHumanN{} accepted candidates & Target-path blocking: \RepairHumanPathN{}/\RepairHumanN{}, blocking with legitimate behavior preserved: \RepairHumanConfirmedN{}/\RepairHumanN{} \\
\bottomrule
\end{tabular}
\par\vspace{4pt}
\noindent\parbox{\columnwidth}{\scriptsize Repair judgments comprise \RepairHumanConfirmedN{} full confirmations, \RepairHumanPartialN{} partial, \RepairHumanIncorrectN{} incorrect and \RepairHumanUnknownN{} unresolved cases. Of the unresolved cases, \RepairHumanUnknownPathN{} block the target path, but additional code context is needed to determine whether legitimate behavior is preserved.}
\end{table}

\section{Security Impact Labeling and Validation}
\label{app:taxonomy-eval}\label{app:codebook}\label{app:taxonomy}\label{app:Ix-construction}

\noindent\textbf{Evidence and class boundaries.}
\label{app:Ix-label-details}\label{app:prompts}
Impact extraction uses cleaned CVE descriptions and records each operation, object, violated requirement and supporting quote.
Mapping applies the fixed codebook to these records without patches or target CWE labels. It preserves distinct evidence-backed violations and leaves insufficient, ambiguous or uncovered statements unclassified.
Deadlock belongs to progress, with a separate coordination violation added when stated.
Progress also includes termination or exception escape beyond an intended isolation boundary.

\noindent\textbf{Counting violation chains.}
A description contributes once to the description count if it contains any cross-class causal link. For the direction analysis, we count each ordered pair of distinct classes attached to successive violations marked as causally linked.
Of \IxChainLinksN{} such links, \IxChainNumMemN{} run from numeric to memory-bounds and \IxChainConStateN{} from concurrency to object-state.
In total, \IxChainUpDownN{} of \IxChainLinksN{} links (\IxChainUpDownShare{}\%) run from numeric, concurrency or input-acceptance to memory-bounds, object-state or reference-and-type; only one runs between these groups in the reverse direction.

\noindent\textbf{Validation.}
The \IxHumanN{}-description audit in Table~\ref{tab:Ap-human-checks} uses cleaned descriptions and class definitions.
Exact class-set agreement includes descriptions with no assigned class, while micro-F1 measures agreement over individual class assignments.

\section{Repair--Impact Associations}
\label{app:assoc}

\begin{table*}[!t]
\centering\footnotesize
\caption{Main repair--impact associations: support and prevalence differences.}
\label{tab:Ap-assoc}
\setlength{\tabcolsep}{4pt}
\renewcommand{\arraystretch}{1.05}
\begin{tabular}{@{}>{\raggedright\arraybackslash}p{.21\textwidth}>{\raggedright\arraybackslash}p{.35\textwidth}rr@{}}
\toprule
Impact class & Repair category / condition & $n_{pi}/n_i$ & $\Delta$ [95\% interval] (pp) \\
\midrule
memory-bounds & Check remaining input & 110/969 & $+$8.8 [4.1, 13.8] \\
 & Check destination capacity & 95/969 & $+$9.5 [6.2, 12.5] \\
 & Check object extent & 214/969 & $+$18.8 [13.0, 24.6] \\
 & Bound processing & 106/969 & $+$8.4 [5.3, 11.8] \\
 & Provide capacity & 81/969 & $+$8.1 [4.6, 12.2] \\
 & Correct size or position & 84/969 & $+$6.2 [3.3, 8.9] \\
\addlinespace[2pt]
object-state & Initialize or clear state & 118/555 & $+$17.2 [5.1, 27.1] \\
 & Protect object lifetime & 208/555 & $+$35.6 [24.2, 50.0] \\
 & Coordinate concurrent access & 50/555 & $+$6.0 [0.9, 10.4] \\
\addlinespace[2pt]
reference-and-type & Check pointer validity & 209/380 & $+$52.1 [43.5, 62.5] \\
\addlinespace[2pt]
numeric & Check numeric validity & 150/290 & $+$47.8 [36.8, 59.5] \\
 & Change numeric representation & 51/290 & $+$16.5 [11.3, 21.9] \\
\addlinespace[2pt]
resource-holding & Release and clean up & 158/233 & $+$65.6 [53.8, 73.4] \\
\addlinespace[2pt]
progress & Check workload & 18/157 & $+$10.6 [2.9, 19.2] \\
\addlinespace[2pt]
concurrency & Coordinate concurrent access & 71/130 & $+$52.9 [14.2, 81.5] \\
\addlinespace[2pt]
input-acceptance & Check state consistency & 11/33 & $+$23.1 [5.4, 42.6] \\
\addlinespace[2pt]
untrusted-resolution & Check permission & 10/65 & $+$13.0 [6.6, 21.4] \\
 & Separate data from instructions & 22/65 & $+$33.7 [21.9, 47.0] \\
 & Restrict privileges or features & 9/65 & $+$12.3 [5.5, 21.2] \\
\addlinespace[2pt]
trust-decision & Check permission & 62/149 & $+$41.1 [15.8, 63.9] \\
 & Restrict privileges or features & 20/149 & $+$12.2 [6.5, 19.8] \\
\addlinespace[2pt]
sensitive-data & Initialize or clear state & 11/35 & $+$24.3 [8.5, 39.2] \\
\bottomrule
\end{tabular}

\end{table*}

The \ChAssocPairedPatches{} dual-labeled patches form \ChFiveBlocks{} repository blocks, with the mainline Linux repository and five development trees grouped together.
Each patch contributes its primary repair category and each assigned impact class.
Eligible repair labels require 20 patches from five blocks. Each tested pair additionally requires 20 patches inside and outside the impact class, with in-class support spanning five blocks.
These rules admit \ChFiveTestable{} pairs but leave insufficient support for testing propagate-failure and crypto-material.

Adjusted prevalence estimates~\cite{agresti2000simple}, repository-block bootstrap intervals and block permutations support inference.
Benjamini--Yekutieli correction~\cite{benjamini2001fdr} applies within each impact class and globally at $q\leq0.10$.
Main links require positive, interval-supported differences and stability across Linux, data-source and edit-size stratifications. All \ChFiveMain{} also pass global correction.
Table~\ref{tab:Ap-assoc} reports the complete set of main links.

\noindent\textbf{Impact agreement within patch clusters.}
Using the \ChUnifiedImpactPatches{} patches with an assigned impact, we count all unordered pairs from different project lineages within each partition's clusters.
A pair agrees if its members share at least one impact class.
The reference is the mean over 300 permutations that keep cluster assignments fixed and shuffle complete impact-label sets among patches in the same repository.

\begin{table}[H]
\centering\footnotesize
\caption{Cross-lineage pairs in the same cluster that share an impact class.}
\label{tab:Ap-cluster-impact-agreement}
\begin{tabular}{@{}lrrr@{}}
\toprule
Partition & Pairs & Observed (\%) & Shuffled (\%) \\
\midrule
CR       & 109,192 & 23.4 & 20.7 \\
Jina     &  90,690 & 41.4 & 28.3 \\
Diff-AST & 282,553 & 24.6 & 21.0 \\
TF-IDF   & 197,558 & 22.7 & 20.7 \\
\bottomrule
\end{tabular}
\end{table}

\section{Repair Conditions and Robustness}
\label{app:repair}

\noindent\textbf{Targets and edit complexity.}
The benchmark contains \ChSixTargets{} held-out C/C++ repair targets with supplied pre-fix code of at most 10,000 characters and at least one reviewed same-mechanism example in the pool.
The code-length criterion is applied to completed runs without resampling examples.
Table~\ref{tab:Ap-complexity} defines the three edit-complexity strata used in Table~\ref{tab:repair-difficulty}.

\begin{table}[H]
\centering\footnotesize
\setlength{\tabcolsep}{3pt}
\caption{Edit-complexity strata.}
\label{tab:Ap-complexity}
\begin{tabular}{@{}>{\raggedright\arraybackslash}p{.15\columnwidth}>{\raggedright\arraybackslash}p{.79\columnwidth}@{}}
\toprule
Stratum & Criteria \\
\midrule
Easy & One function, at most one hunk and six changed lines, and no new identifiers. \\
Medium & Targets outside the easy and hard strata. \\
Hard & At least two functions, more than 15 changed lines, at least two new identifiers, or at least four hunks. \\
\bottomrule
\end{tabular}
\par\vspace{4pt}
\noindent\parbox{\columnwidth}{\scriptsize New identifiers occur in added lines but not in the pre-fix functions, excluding keywords.}
\end{table}

\noindent\textbf{Retrieval.}
The pool provides function-level examples from \ChSixPoolPatches{} patches.
Each query excludes its target implementation, linked CVEs, fixing commits and repair family. Same-lineage examples remain eligible in the main comparison.
The frozen reducer and nearest centroid assign additional patches to the existing clusters.
For query function changes $d_f$ and example vector $V_u$, ranking uses $S(u)=\max_f\cos(d_f,V_u)$.
Selection takes distinct repair families, prefers distinct repository families and fills from neighboring clusters and then the global pool.
Impact-guided retains the top three repair categories under the target impact, using family-counted conditional frequencies with global-frequency smoothing and equal weighting of multiple impacts.
Its category filter remains active during fallback.
Missing impact labels or fewer than three eligible examples use Zero-shot.

Each generator supplies three hypothetical patches, each defining a query group, with developer changed-line or insertion-position hints and no replacement code.
Unusable query groups use Zero-shot.
Oracle uses developer-informed reviews, prioritizing same-mechanism examples and filling remaining slots with examples sharing necessary repair elements.
Table~\ref{tab:Ap-query-rates} reports pass@3 for developer and model-generated queries.

\begin{table}[H]
\centering\footnotesize
\setlength{\tabcolsep}{4pt}
\caption{Mean pass@3 (\%) by query type across all \ChSixTargets{} targets, including Zero-shot fallbacks.}
\label{tab:Ap-query-rates}
\begin{tabular}{@{}lrr@{}}
\toprule
Method & Developer fix & Model-generated \\
\midrule
Cluster-guided & \ChSixMeanCluster{} & \ChSixMeanClusterHyp{} \\
Impact-guided & \ChSixMeanImpact{} & \ChSixMeanImpactHyp{} \\
\bottomrule
\end{tabular}
\par\vspace{4pt}
\noindent\parbox{\columnwidth}{\scriptsize Zero-shot and Random average \ChSixMeanZero{}\% and \ChSixMeanRandom{}\%, respectively. Model-generated queries receive change-location hints.}
\end{table}

\noindent\textbf{Assessment and aggregation.}
The first generation turn requests a repair plan, followed by corrected functions in the second turn.
Unchanged, unparseable, failed or truncated outputs fail.
Exact normalized developer-diff matches are accepted. Other candidates receive source-level assessment using the vulnerable code, reference diff and candidate diff, with generator, selection method and examples withheld.
Acceptance requires blocking the reference dangerous path with the same control-flow effect, including equivalent implementations of that effect.
Candidates are assessed from source without compilation or execution.
In the human review of \RepairHumanN{} accepted candidates, \RepairHumanPathN{} blocked the target path and \RepairHumanConfirmedN{} also preserved legitimate behavior (Table~\ref{tab:Ap-human-checks}).

For each target, a group scores one if any of its three candidates passes and zero otherwise.
We average the three group scores for Random and for retrieval with model-generated queries. The other conditions have one group per target. The reported pass@3 is the mean of these target scores, including fallbacks, with every target weighted equally.
The lineage ablation excludes the target's recorded lineage, and the Jina ablation replaces both pool and query representations.
Table~\ref{tab:repair-ablation} reports paired differences for these comparisons.

\begin{table}[H]
\centering\footnotesize
\caption{Mean paired ablation differences across four models (pp).}
\label{tab:repair-ablation}
\begin{tabular}{@{}lrr@{}}
\toprule
Contrast & Targets & $\Delta$ (pp) \\
\midrule
\multicolumn{3}{@{}l}{\emph{Developer minus model-generated query}} \\
Cluster-guided & 285 & \ChSixQueryGainCluster{} \\
Impact-guided & 285 & \ChSixQueryGainImpact{} \\
\midrule
\multicolumn{3}{@{}l}{\emph{Impact-guided minus Cluster-guided}} \\
developer query & \ChSixTargets{} & \ChSixFilterDev{} \\
model-generated query & \ChSixTargets{} & \ChSixFilterHyp{} \\
Gap difference (developer minus model query) & \ChSixTargets{} & \ChSixFilterDiff{} \\
\midrule
\multicolumn{3}{@{}l}{\emph{Same-lineage examples: allowed minus excluded}} \\
Random & \ChSixTargets{} & $-0.3$ \\
Cluster-guided & \ChSixTargets{} & $+3.5$ \\
Impact-guided & \ChSixTargets{} & $+3.0$ \\
Impact-guided (model query) & \ChSixTargets{} & $+1.7$ \\
Oracle & \ChSixLineageOracleN{} & $+3.4$ \\
\midrule
\multicolumn{3}{@{}l}{\emph{Cross-lineage: Impact-guided minus Random}} \\
gain, developer query & \ChSixTargets{} & \ChSixXplcImpactDevGain{} \\
change vs.\ shared-lineage pool & \ChSixTargets{} & \ChSixXplcImpactDevChange{} \\
gain, model-generated query & \ChSixTargets{} & \ChSixXplcImpactGain{} \\
change vs.\ shared-lineage pool & \ChSixTargets{} & \ChSixXplcImpactChange{} \\
\midrule
\multicolumn{3}{@{}l}{\emph{Jina minus CR}} \\
Cluster-guided & \ChSixTargets{} & \ChSixJinaCluster{} \\
Impact-guided & \ChSixTargets{} & \ChSixJinaImpact{} \\
Cluster-guided (model query) & \ChSixTargets{} & \ChSixJinaClusterHyp{} \\
Impact-guided (model query) & \ChSixTargets{} & \ChSixJinaImpactHyp{} \\
\midrule
\multicolumn{3}{@{}l}{\emph{Impact filter under Jina}} \\
model-generated query & \ChSixTargets{} & \ChSixJinaFilterHyp{} \\
\bottomrule
\end{tabular}

\par\vspace{3pt}
\noindent\parbox{\columnwidth}{\scriptsize The first two rows omit unusable model-query groups and average over targets with paired scores for all four models. The other comparisons include Zero-shot fallbacks. All differences use unrounded scores.}
\end{table}

\noindent\textbf{Sensitivity to the evaluator.}
On the 218 targets with overlapping evaluations, requiring both gpt-oss-120b and Qwen3.8-27B to accept a candidate preserved the gains over Random for all four generators under Cluster-guided retrieval with a developer query and Impact-guided retrieval with a model-generated query.
For fixes generated by GPT-6 Luna, these gains were 9.8 and 7.8 percentage points under the two-model rule, compared with 9.5 and 7.2 under Luna's own judgments. Each open-weight evaluator also found both gains when scored separately.

\noindent\textbf{Repair plans.}
Plan assessment uses a fixed reference for the unsafe operation, violated condition and required repair effects.
The generator identity, selection method and final candidate are withheld during plan assessment.
Problem identification requires the matched diagnosis, while completeness requires all repair elements.
Scores average attempts within targets, then targets and generators.

\end{document}